\documentclass[reprint,eqsecnum,floats,aps,amsmath,amssymb,nofootinbib,prd,onecolumn,showpacs]{revtex4-2}
\usepackage{bm}
\usepackage{graphicx,physics}
\usepackage{amsmath,amssymb,mathtools,mathrsfs}
\usepackage{hyperref}
\usepackage{graphicx}
\usepackage{arydshln}
\usepackage{xcolor}
\usepackage{braket}
\usepackage{tensor}
\usepackage{enumitem,array,textcomp}

\begin{document}

\title{Axial perturbations in Kantowski-Sachs spacetimes and hybrid quantum cosmology}

\author{Guillermo A. Mena Marug\'an}
\email{mena@iem.cfmac.csic.es}
\affiliation{Instituto de Estructura de la Materia, IEM-CSIC, C/ Serrano 121, 28006 Madrid, Spain}
\author{Andr\'es M\'{\i}nguez-S\'anchez}
\email{andres.minguez@iem.cfmac.csic.es}
\affiliation{Instituto de Estructura de la Materia, IEM-CSIC, C/ Serrano 121, 28006 Madrid, Spain}

\begin{abstract}
Recently, there has been a growing interest in investigating homogeneous but anisotropic spacetimes owing to their relation with nonrotating, uncharged black hole interiors. We present a description of axial perturbations for a massless scalar field minimally coupled to this geometry. We truncate the action at the quadratic perturbative order and tailor our analysis to compact spatial sections. Perturbations are described in terms of perturbative gauge invariants, linear perturbative constraints, and their canonically conjugate variables. The entire set, encompassing perturbations and homogeneous degrees of freedom, is consolidated into a canonical one. We employ a hybrid approach to quantize this system, integrating a quantum representation of the homogeneous sector using loop quantum cosmology techniques with a conventional field quantization of the perturbations.
\end{abstract}

\maketitle

\section{Introduction \label{sec: I}}

The foundation of general relativity (GR) provides a robust framework for understanding gravity as the curvature of spacetime caused by the presence of mass and energy. However, obtaining exact solutions to Einstein's equations is challenging or even impossible in the majority of gravitational scenarios. It is under such circumstances that perturbation theory comes into play. Its significance lies not only in its utility for solving practical problems, but also in its capacity to unveil new aspects of the universe. By studying small fluctuations around well-known solutions, valuable insights can be gained into the stability of astrophysical systems, gravitational radiation, the formation and evolution of black holes, and other cosmic phenomena. Its practical applications extend to fields like observational astrophysics and the detection of gravitational waves. For instance, the observation of merging binary black hole systems, using interferometers, significantly benefits from perturbative methods to model and understand events which generate these phenomena, i.e. ringdown processes \cite{ringdown}.

A challenge in perturbation theory within GR, stemming from the diffeomorphism invariance of the theory, is addressing the degrees of freedom that characterize perturbations through the use of perturbative gauge invariant variables \cite{Bardeen,Mukhanov,Sasaki,RW}. We recall that gauge invariant quantities do not change under gauge transformations; in our case, this implies that perturbative gauge invariant variables do not depend on how we relate our well-known spacetime, commonly referred to as background, to our perturbed spacetime. Therefore, the choice of these variables is essential to ensure the consistency and validity of the results. Under this paradigm, the Hamiltonian formalism of GR provides an intuitive description for identifying and characterizing perturbative gauge invariant variables at any perturbative order, as addressed in a pioneering work in Ref. \cite{M}. Although the complete formulation of the theory would involve considering all perturbative orders, full calculations can often become computationally intensive and, in some cases, even infeasible. Hence, it is customary in most cases to work up to the leading perturbative order, assuming that higher-order corrections are negligible. Employing this approach not only simplifies computational aspects but also preserves the most fundamental properties of the problem, thereby enhancing the ease of analysis and interpretation of the results.

Unraveling the quantum nature of gravity is one of the fundamental challenges in modern physics. Despite the formidable task of the scientific community in understanding our surroundings in the universe, certain physical regimes persistently elude complete comprehension. Among gravitational systems, notable examples are the early Universe and black holes. It is anticipated that a theory adept at integrating the principles of quantum mechanics with the foundations of GR will provide insights into the understanding of these scenarios. One of the most robust theories in contemporary physics to achieve this goal is loop quantum gravity (LQG). Essentially, LQG is characterized as a nonperturbative, diffeomorphism invariant, and background-independent quantization of GR. Moreover, this theory offers a distinctive perspective by depicting Einstein's theory using a $SU(2)$ connection and its densitized triad as the canonical set of variables within the Hamiltonian framework. These variables give rise to the formation of holonomies over closed paths (loops) and fluxes of the densitized triad over closed surfaces, constituting the algebra of functions that is represented quantum mechanically \cite{A&L, Thiemann}. The application of LQG techniques to the study of simplified systems is a specialized discipline known in the literature as loop quantum cosmology (LQC) \cite{A&S,GMM}. One of the groundbreaking achievements of LQC is the resolution of the initial singularity in standard cosmology, accomplished by avoiding it through a bouncing phenomenon known as the big bounce \cite{APS1,APS2}.

In gravitational research, the Kantowski-Sachs spacetime has recently gained significant attention, standing out as a particularly interesting model to study. Its geometry defines a homogeneous but anisotropic scenario, as outlined in Refs. \cite{K&S, E.Weber}, but it also holds profound significance owing to its diffeomorphic equivalence to the interior of a nonrotating, uncharged black hole. To unveil this relationship, we can conduct a compelling comparison between Kantowski-Sachs spacetime and Schwarzschild. Despite both spacetimes featuring spherical symmetry, a noteworthy distinction arises, namely, Kantowski-Sachs features a radial Killing field, whereas the exterior Schwarzschild exhibits a temporal one. In Schwarzschild, the transition from the exterior to the interior geometry involves an inversion of radial and temporal coordinates. This property, in turn, leads us to the relationship we sought between the Schwarzschild interior and Kantowski-Sachs. Regarding the quantum nature of Kantowski-Sachs with an eye to black hole physics, the LQC formalism initially did not attain as significant progress as its counterpart in standard cosmology. A (nonexhaustive) list of works in this field includes the following Refs. \cite{A&B,LBH1,LBH2,B&V,Chiou,LBH3,LBH4,LBH5,JPS,LBH6,LBH7,LBH8,LBH9,LBH10,LBH11,LBH12}. Nevertheless, this area of research has experienced a revitalization, driven by a series of recent publications that showcase promising advancements beyond the initial works \cite{AOS,AOS2,AO,BH_Con,GQM2,BH_GAB,BH_Cong,GBA}. In fact, a comprehensive quantization of this spacetime has been successfully achieved using the same loops methods that were employed in the quantization of standard cosmology. The only distinctive element in the quantization program, diverging from prior LQC works, is the necessity to contemplate an extended Hamiltonian formulation of the problem to accomplish quantization \cite{BH_GAB}. Remarkably, this extension poses no apparent obstruction; instead, it facilitates and streamlines the quantization process \cite{GBA}. This outcome represents a significant breakthrough as, before these works, only an effective description incorporating quantum corrections from LQC was available. Moreover, it provides the groundwork for a variety of studies that can delve into the nature and properties of this quantum spacetime.

Taking into account all these comments, the objective of this paper is twofold: To employ a perturbative description of Kantowski-Sachs spacetimes and to improve our understanding of the geometry quantization by introducing a quantum formalism specifically designed to handle these perturbations. For the sake of completeness, our study incorporates the coupling of the geometry with a homogeneous, massless scalar field. A good starting point to initiate the discussion is the analysis presented in Ref. \cite{DC}, dedicated to the study of perturbations in spherically symmetric spacetimes. For this scenario, Kantowski-Sachs serves as a particular case of background spacetime, while the perturbative description is notably streamlined by considering the symmetries of Kantowski-Sachs. In fact, symmetries enable us to decompose perturbations in terms of spherical harmonics. Given its importance in the investigation of gravitational radiation, we employ the Regge-Wheeler-Zerilli basis for these harmonics. When conducting the decomposition, special attention is paid to the behavior of harmonics under parity transformations in angular coordinates. This behavior allows us to divide our study into two independent cases: The axial and the polar cases. For the scope of this article, we specifically delve into the study of axial modes, chosen for their computational simplicity. Taking advantage of the additional symmetry present in Kantowski-Sachs compared to generic spherically symmetric spacetimes, our work proposes to complement the harmonic decomposition with a Fourier decomposition (in the radial direction), significantly simplifying the calculations. By employing a series of canonical transformations, dependent on both perturbative modes eigenvalues and background variables, and incorporating various redefinitions of the Lagrange multipliers, not previously considered for this spacetime, we will show how to derive a Hamiltonian formulation. It is worth mentioning that the validity of these transformations and redefinitions holds at the considered perturbative order. The resulting framework features canonical variables for the perturbations that are either perturbative gauge invariant variables or perturbative gauge degrees of freedom, the latter being decoupled from the former and not pertinent to the ensuing discussion \cite{Lan,DC,LMM}. 

After these transformations are completed, the dynamics are governed by a diagonal Hamiltonian, dependent solely on variables that maintain perturbative gauge invariance at leading order. Furthermore, its expression is entirely adapted to the results obtained in Ref. \cite{AT}. In that article, under relatively nonrestrictive conditions, it is proved that the Fock quantization of a free scalar field within the Kantowski-Sachs geometry admits a privileged, unitary equivalent family of representations. The extension to our scenario, considering Kantowski-Sachs perturbations, is straightforward and suggests a promising direction for the quantization of the perturbations. The integration of the quantum description of the perturbed Kantowski-Sachs spacetime is conducted using a formalism well known in the literature as hybrid loop quantum cosmology (LQC-Hybrid). A detailed analysis of this method can be found in Ref. \cite{hyb-review}, where its successful application in cosmology is reviewed \cite{FMO,hyb1,hyb2}. For the first time, these techniques are being applied to Kantowski-Sachs. In brief, the LQC-Hybrid approach involves utilizing a well-established quantum cosmology theory for the treatment of the background, while assigning perturbations a less dominant role in quantum geometry effects. This allows for their treatment through a more conventional representation. For our discussion, we will employ the LQC representation for the background and a Fock (or a Schrödinger) representation for the perturbations. Using this framework, both the background and the perturbations are addressed altogether through pure quantum mechanical means. This contrasts with another approach frequently employed in the LQC community, namely the dressed metric formalism \cite{AAN,AAN2,AAN3}, in which perturbations are effectively treated as test fields in a background dressed with quantum modifications.

The article is organized as follows. In Sec. \ref{sec: II}, we introduce the fundamental concepts of the Hamiltonian formulation within GR, covering both perturbed and unperturbed scenarios. Section \ref{sec: III} presents a concise overview of a homogeneous but anisotropic spacetime coupled with a massless scalar field. The discussion employs canonical variables based on the connection formulation of LQG. Moving forward in Sec. \ref{sec: IV} we introduce a basis of functions motivated by the symmetries of the background spacetime that facilitates an appropriate description of the perturbations in Kantowski-Sachs. The study specializes in axial perturbations in Sec. \ref{sec: V}, providing a detailed formulation in terms of perturbative gauge invariant variables. The approach we adopt maintains the entire canonical structure of the system, encompassing both the background and the perturbations. Section \ref{sec: VI} builds upon the results from previous sections to conduct a quantization of the system. Finally, in Sec. \ref{sec: VII}, we present and discuss the implications drawn from our results. To address the subtleties concerning the global canonical structure, that are not relevant for the main discussion, we include an appendix. Throughout this article, we adopt geometric natural units, setting the speed of light, the Planck constant, and the gravitational constant to unity.

\section{Hamiltonian formalism for perturbations \label{sec: II}}

In this section, we will briefly revisit the Hamiltonian formulation of GR coupled with a massless scalar field and explore its perturbations. Although this description has been previously addressed in the literature, and we refer to Ref. \cite{Thiemann} for a detailed discussion of the unperturbed case and to Ref. \cite{DC} (or, alternatively, to Refs. \cite{DB3,DB2,DB1}) for the perturbative treatment, we will provide a concise review to establish the notation for our upcoming sections.

The starting point of the unperturbed formalism is the Einstein-Hilbert action coupled with a massless scalar field $\Phi$. For any spacetime that allows an Arnowitt-Deser-Misner (ADM) decomposition, a time vector field, referred to as $\partial_t$, exists. This vector field serves to foliate the spacetime into a collection of spatial Cauchy hypersurfaces. With the aim of simplifying the discussion, we introduce coordinates $x^{\mu} = (t,x^a)$ aligned with this foliation, where the label $\mu$ represents spacetime indices, and $a$ stands for spatial indices. Consequently, the action can be written as
\begin{equation}
    S_0 = \frac{1}{\kappa} \int_{\mathbb{R}}\text{d}t\int_{\sigma}\text{d}^3x \left( \dot{\Phi}\Pi_{\Phi} + \dot{g}_{ab}\Pi^{ab} - N\mathcal{H} - N^a\mathcal{H}_a \right).
\end{equation}
Here, $\kappa$ is $16\pi$ in the chosen units, and the integration is performed over the globally hyperbolic manifold $\mathbb{R}\times\sigma$, where $\sigma$ is the spatial manifold. The notation $\dot{g}_{ab}$ and $\dot{\Phi}$ refer to the Lie derivative of the spatial metric and the scalar field, respectively, along the flow generated by $\partial_t$. In our coordinates, both can be expressed as $g_{ab,t}$ and $\Phi_{,t}$, where the comma denotes partial derivative. Simultaneously, the expressions for their canonical conjugate momenta are
\begin{equation}
    \Pi^{ab} = \kappa \frac{\delta S_0}{\delta \dot{g}_{ab}} = \sqrt{\mathfrak{g}}(K^{ab} - g^{ab}\tensor{K}{^c_c}), \qquad \Pi_{\Phi} = \kappa \frac{\delta S_0}{\delta \dot{\Phi}} = \frac{\sqrt{\mathfrak{g}}}{N}(\Phi_{,t} - N^a\Phi_{,a}), \qquad \mathfrak{g} = \text{det}(g_{ab}).
\end{equation}
In these formulas, we introduced the lapse function $N$, the shift vector $N^a$, and the extrinsic curvature $K_{ab}$. Lastly, the expressions for the Hamiltonian constraint $\mathcal{H}$ and the spatial diffeomorphism constraints $\mathcal{H}_a$ are
\begin{equation}
    \begin{aligned}
        &\mathcal{H} = \frac{1}{\sqrt{\mathfrak{g}}} \left[ g_{ac}g_{bd} - \frac{1}{2}g_{ab}g_{cd} \right] \Pi^{ab}\Pi^{cd} - \sqrt{\mathfrak{g}}R + \frac{1}{2}\left[\frac{\Pi_{\Phi}^2}{\sqrt{\mathfrak{g}}}+\sqrt{\mathfrak{g}}g^{ab}\Phi_{,a}\Phi_{,b}\right]=0,\\ 
        &\mathcal{H}_a = -2g_{ac}D_b\Pi^{bc} + \Pi_{\Phi}\Phi_{,a}=0,
    \end{aligned}
\end{equation}
where $D$ represents the covariant derivative with respect to the spatial metric, satisfying the condition $D_{a}g_{bc}=0$, and $R$ is the Ricci scalar for the spatial sections.

To carry out the perturbative analysis, we introduce a dimensionless parameter, denoted as $\epsilon$, serving a key role in establishing the hierarchy of perturbations and labeling a family of manifolds $\mathcal{M}(\epsilon)$ within our system. When $\epsilon = 0$, the manifold corresponds to the background, and for all other cases, it represents a perturbed manifold. Our aim is to compare perturbative quantities with background quantities while anticipating the presence of gauge freedom associated with the choice of the mapping, $\psi_{\epsilon}:\mathcal{M}(0)\rightarrow\mathcal{M}(\epsilon)$, between these manifolds. Hence, when dealing with a perturbative quantity $\Tilde{g}(\epsilon)$ on $\mathcal{M}(\epsilon)$, its pullback $\psi_{\epsilon}^*\Tilde{g}(\epsilon)$ defined on $\mathcal{M}(0)$ unfolds as follows \cite{DC,DB2}:
\begin{equation}
     \psi_{\epsilon}^*\Tilde{g}(\epsilon) = g + \sum_{n=1}^{\infty}\frac{\epsilon^n}{n!}\Delta_{\psi}^n[g], \qquad \Delta_{\psi}^n[g] = \left.\frac{\text{d}^n\psi_{\epsilon}^*\Tilde{g}(\epsilon)}{\text{d}\epsilon^n}\right|_{\epsilon=0}.
\end{equation}
The inclusion of the label $\psi$ in the perturbations, which we will later omit for simplicity, serves as a reminder that, in general, establishing a map is essential for accurately defining these perturbations. With  this terminology in mind, our focus shifts to examining leading-order perturbations, elucidated by the effective second-order action, denoted as
\begin{equation}
    \label{eq: II1}
    \frac{1}{2}\Delta^2_1[S_0] = \frac{1}{\kappa} \int_{\mathbb{R}}\text{d}t\int_{\sigma}\text{d}^3x \left( h_{ab,t}p^{ab} + \varphi_{,t}p - C\Delta[\mathcal{H}] - B^a\Delta[\mathcal{H}_a] - \frac{N}{2}\Delta^2_1[\mathcal{H}] - \frac{N^a}{2}\Delta^2_1[\mathcal{H}_a] \right).
\end{equation}
This equation incorporates the following notation, introduced to handle first-order perturbative contributions:
\begin{equation}\label{notati}
    C = \Delta[N], \qquad B^a = \Delta[N^a], \qquad h_{ab} = \Delta[g_{ab}], \qquad p^{ab} = \Delta[\Pi^{ab}], \qquad \varphi = \Delta[\Phi], \qquad p = \Delta[\Pi_{\Phi}].
\end{equation} 
The lack of a first-order action term to address leading-order perturbations is grounded in its vanishing contribution when dealing with compact spatial sections, as will become apparent later in our model. For similar reasons, the contributions of second-order perturbations of the metric and the scalar field are irrelevant in our present discussion. To emphasize this point, we adopted the notation $\Delta^2_1$ for second-order quantities that are composed of first-order perturbations. To conclude, the terms $\Delta[\mathcal{H}]$ and $\Delta[\mathcal{H}_a]$ should be understood as the first-order perturbative gauge constraints, while $\Delta_1^2[\mathcal{H}]$ and $\Delta_1^2[\mathcal{H}_a]$ constitute the dominant perturbative contributions to the (originally unperturbed) total Hamiltonian. These terms determine the dynamics of the perturbations, and their expressions described using the perturbative variables introduced in Eq. \eqref{notati} can be found in Chap. 4 of Ref. \cite{DC} or in Ref. \cite{DB3}.

\section{Kantowski-Sachs spacetimes \label{sec: III}}

From now on, our attention will turn exclusively to Kantowski-Sachs, a homogeneous but anisotropic spacetime. For convenience and simplicity in our calculations, we adopt compact spatial sections $\sigma_o = S_o^1 \times S^2$, where $S_o^1$ is a 1-sphere, with a period of $L_o$, and $S^2$ represents the standard 2-sphere. As we will see, this compactification will allow us to isolate the zero modes of the perturbed system. The noncompactified limit should be reached in a suitable limit with large $L_o$. Under these considerations, we present the ADM splitting of the Kantowski-Sachs metric as
\begin{equation}
    \text{d}s^2 = - N^2(t) \text{d}t^2 + a^2(t) \text{d}x^2 + r^2(t) \text{d}\Omega^2, 
\end{equation}
where $a(t)$ and $r(t)$ are two scale factors that determine the 3-metric, and $\text{d}\Omega^2$ is the line element for the 2-sphere. Conversely, the conjugate momentum of the spatial metric is dictated by $\Tilde{\Pi}^{xx}(t)$ and $\Tilde{\Pi}^{\theta\theta}(t)$. The tilde notation is used to indicate that these quantities have been density-unweighted with respect to the 2-sphere metric. To align our discussion with the variables used in other LQC works \cite{A&B, AOS}, we introduce the following canonical transformation:
\begin{equation}
    a^2 = \frac{p_b^2}{L_o^2|p_c|}, \qquad r^2 = |p_c|, \qquad \Tilde{\Pi}^{xx} = -\frac{2 L_o b|p_c|}{\gamma p_b}, \qquad \Tilde{\Pi}^{\theta\theta} = -\frac{1}{\gamma L_o |p_c|} \left(cp_c + b p_b\right),
\end{equation}
which establish a connection between metric variables and their momenta, on the one hand, and connection variables $\{b, c\}$ and their momenta, on the other hand. Despite working with metric variables is feasible, this canonical transformation constitutes a crucial step for the next sections. The Kantowski-Sachs phase space can be described using these variables. After a phase space reduction via a spatial integration over $\sigma_o$, the contribution of the geometric degrees of freedom to the symplectic two-form is
\begin{equation}
    \Theta^{\text{B}} = \frac{1}{\gamma}\text{d}b\wedge\text{d}p_b + \frac{1}{2\gamma}\text{d}c\wedge\text{d}p_c, \qquad \{b,p_b\}_{\text{B}} =  \gamma, \qquad \{c,p_c\}_{\text{B}} = 2\gamma,
\end{equation}
while the contribution of the scalar field and its redefined momentum $\Tilde{\Pi}_{\Phi}=L_o \Pi_{\Phi}/(4 \sin\theta)$ is
\begin{equation}
     \Theta^{\Phi} = \text{d}\Phi\wedge\text{d}\Tilde{\Pi}_{\Phi}, \qquad \{\Phi,\Tilde{\Pi}_{\Phi}\}_{\Phi} = 1.
\end{equation}

Homogeneity leads us to assume that $\Phi$ depends solely on time and further eliminates the diffeomorphism constraints. The gauge freedom associated with $\mathcal{H}_a$ is effectively removed by setting $N^a$ to zero. As a result, in the Hamiltonian theory, only the Hamiltonian constraint remains, and upon spatial integration, it can be expressed as
\begin{equation}
    \label{eq: III1}
    \Tilde{H}_{\text{KS}}[\Tilde{N}] = -\Tilde{N}L_o\left[ \Omega_b^2 + \frac{p_b^2}{L_o^2} + 2\Omega_b\Omega_c - 4\frac{\Tilde{\Pi}_{\Phi}^2}{L_o^2}\right], \qquad \Omega_j = \frac{jp_j}{\gamma L_o}\; \text{for}\; j=b\;\text{or}\;c.
\end{equation}
For instance, for $j=b$ we have $\Omega_b=bp_b/(\gamma L_0)$. The lapse function is defined as $\Tilde{N} = 2\pi NL_o/\mathcal{V}$, with $\mathcal{V}$ the volume of the spatial sections, to ensure consistency with Ref. \cite{GBA} (up to an inconsequential global factor equal to $L_o^2$). This redefinition assigns a density weight of $-1$ to $\Tilde{N}$, in accordance with the convention commonly employed in LQG, and a density weight of $2$ to the Hamiltonian $\Tilde{H}_{\text{KS}}$, to balance the overall expression. The background action, after a spatial integration, becomes
\begin{equation}
    S_0 = \int_{\mathbb{R}} \text{d}t \left( -\frac{1}{2\gamma}(p_c)_{,t}c - \frac{1}{\gamma}(p_b)_{,t}b + \Phi_{,t}\Tilde{\Pi}_{\Phi} - \Tilde{H}_{\text{KS}}[\Tilde{N}] \right).
\end{equation}
The equations of motion for the phase space variables, once the constraints are imposed, are given by
\begin{equation}
    \frac{\delta S_0}{\delta b} = -\frac{1}{\gamma}(p_b)_{,t} + 2\Tilde{N}\frac{p_b}{\gamma}(\Omega_b + \Omega_c) = 0, \quad \frac{\delta S_0}{\delta c} = -\frac{1}{2\gamma}(p_c)_{,t} + 2\Tilde{N}\frac{p_c}{\gamma}\Omega_b = 0, \quad \frac{\delta S_0}{\delta \Phi} = - (\Tilde{\Pi}_{\Phi})_{,t} = 0,
\end{equation}
\begin{equation}
    \frac{\delta S_0}{\delta p_b} = \frac{1}{\gamma}b_{,t} + 2\Tilde{N}\frac{b}{\gamma}\left( \Omega_b + \Omega_c + \frac{\gamma p_b}{bL_o} \right) = 0, \quad \frac{\delta S_0}{\delta p_c} = \frac{1}{2\gamma}c_{,t} + 2\Tilde{N}\frac{c}{\gamma}\Omega_b = 0, \quad \frac{\delta S_0}{\delta \Tilde{\Pi}_{\Phi}} = \Phi_{,t} - 8\Tilde{N}\Tilde{\Pi}_{\Phi} = 0.
\end{equation}
The time evolution of any function $f$ defined on the reduced phase space can be calculated as $f_{,t} = \{f,H_{\text{KS}}[\Tilde{N}]\}$, where the Poisson brackets refer to $\Theta = \Theta^{\text{B}} + \Theta^{\Phi}$. Indeed, the results presented in Refs. \cite{A&B, AOS} are recovered when $\Phi$ is made equal to zero and the inverse of $\Tilde{N}$ is set to $2\Omega_b$. When associating this spacetime with the interior of a Schwarzschild black hole, the quantity $\Omega_c$, which is a constant of motion, corresponds to the ADM mass value.

\section{Spherical harmonics and Fourier modes \label{sec: IV}}

Kantowski-Sachs spacetime exhibits spherical symmetry and in addition possesses the Killing vector field $\partial_x$, which allows us to carry out a specific treatment of the perturbations. To take full advantage of this fact, it is worthwhile adopting the notation recommended in Refs. \cite{DC,DB1}. This is the objective of this section.

When dealing with perturbations in a spherically symmetric background, the expansion in spherical harmonics proves highly helpful. Our discussion remarks the importance of the Regge-Wheeler-Zerilli basis, which proves very convenient e.g. in the analysis of gravitational radiation. In this basis, any function $\zeta$ on the 2-sphere can be expressed by means of the expansion
\begin{equation}
    \zeta(\theta,\phi) = \sum_{l=0}^{\infty}\sum_{m=-l}^{l}\zeta_l^m Y_l^m, \qquad \zeta_l^m = (Y_l^m,\zeta) = \int_{S^2}\text{d}\Omega \: Y_l^{m*}\zeta, \qquad (Y_{l'}^{m'},Y_l^m) = \int_{S^2}\text{d}\Omega \: Y_{l'}^{m'*}Y_l^m = \delta_{ll'}\delta_{mm'},
\end{equation}
where the symbol $*$ denotes complex conjugation and the spherical harmonics $Y_l^m(\theta,\phi)$ are defined so that
\begin{equation}
    \begin{aligned}
        &\gamma^{AB}\mathcal{D}_{A}\mathcal{D}_{B} Y_l^m = -l(l+1)Y_l^m \quad \text{for} \quad l=0,1,2,\cdots, \; m=-l,\cdots,l,\\
        &Y_l^m(\theta,\phi) \xrightarrow{\textbf{P}} Y_l^m(\pi-\theta,\pi+\phi) = (-1)^lY_l^m(\theta,\phi).
    \end{aligned}
\end{equation}
In these formulas, capital letters from the beginning of the Latin alphabet denote $S^2$ indices, \textbf{P} is a parity transformation, and $\mathcal{D}$ represents the covariant derivative associated with the 2-sphere metric $\gamma_{AB}$. We thus have $\mathcal{D_C}\gamma_{AB} := \gamma_{AB:C} = 0$. 

We now introduce additional harmonics beyond scalars. For them, it is advantageous to differentiate between harmonics with parity $(-1)^l$, referred to as \textit{polar}, and those with parity $(-1)^{l+1}$, known as \textit{axial}. According to this definition, it is obvious that the spherical harmonics $Y_l^m$ are polar. The main reason to separate modes in this way lies in the absence of coupling between them at leading perturbative order. Extending the previous analysis to any covector $w_{A}$ defined on the cotangent space of $S^2$, we can expand it using the orthogonal basis $\{\tensor{Z}{_l^m_A},\tensor{X}{_l^m_{A}}\}$,
\begin{equation}
    w_{A}(\theta,\phi) = \sum_{l=1}^{\infty}\sum_{m=-l}^{l}(\mathcal{W}_l^m \tensor{Z}{_l^m_A} + w_l^m \tensor{X}{_l^m_{A}}), \qquad \mathcal{W}_l^m = \frac{1}{l(l+1)}(\tensor{Z}{_l^m_A},w^A), \qquad w_l^m = \frac{1}{l(l+1)}(\tensor{X}{_l^m_A},w^A), 
\end{equation}
where the harmonics $\tensor{Z}{_l^m_A} = \tensor{Y}{_l^m_{:A}}$ have polar parity, whereas $\tensor{X}{_l^m_A} = \epsilon_{AB}\gamma^{BC}\tensor{Y}{_l^m_{:C}}$ are axial, with $\epsilon_{AB}$ being the Levi-Civita tensor. Both basis elements are null when $l$ is equal to zero, and they are normalized in such a way that $(\tensor{Z}{_{l'}^{m'}_A},\tensor{Z}{_l^m^A}) = (\tensor{X}{_{l'}^{m'}_A},\tensor{X}{_l^m^A}) = l(l+1) \delta_{ll'}\delta_{mm'}$. Similarly, any symmetric 2-tensor $T_{AB}$ can be expanded using the orthogonal basis $\{ \tensor{X}{_l^m_{AB}}, Y_l^m\gamma_{AB}, \tensor{Z}{_l^m_{AB}} \}$,
\begin{equation}
    T_{AB}(\theta,\phi) = \sum_{l=0}^{\infty}\sum_{m=-l}^{l}\Tilde{T}_l^m\gamma_{AB}Y_l^m + \sum_{l=2}^{\infty}\sum_{m=-l}^{l}(T_l^m\tensor{Z}{_l^m_{AB}} + t_l^m\tensor{X}{_l^m_{AB}}),
\end{equation}
where the harmonics $\tensor{X}{_l^m_{AB}}$ and $\tensor{Z}{_l^m_{AB}}$ are trace-free, and become zero when $l$ is equal to zero or to one. Their expressions and normalization in terms of spherical harmonics are
\begin{equation}
    \begin{aligned}
        \tensor{X}{_l^m_{AB}}& = \frac{1}{2}(\tensor{X}{_l^m_{A:B}} + \tensor{X}{_l^m_{B:A}}), \qquad \tensor{Z}{_l^m_{AB}} = \tensor{Y}{_l^m_{:AB}} + \frac{l(l+1)}{2}\gamma_{AB}Y_l^m,\\
        &(\tensor{X}{_{l'}^{m'}_{AB}},\tensor{X}{_l^m^{AB}}) = (\tensor{Z}{_{l'}^{m'}_{AB}},\tensor{Z}{_l^m^{AB}}) = \frac{1}{2}\frac{(l+2)!}{(l-2)!}\delta_{ll'}\delta_{mm'}.
    \end{aligned}
\end{equation}
Conversely, the coefficients of the expansion can be calculated as
\begin{equation}
    \Tilde{T}_l^m = \frac{1}{2} (Y_l^{m},\tensor{T}{^A_A}), \qquad T_l^m = 2\frac{(l-2)!}{(l+2)!} (\tensor{Z}{_l^m_{AB}},T^{AB}), \qquad t_l^m = 2\frac{(l-2)!}{(l+2)!} (\tensor{X}{_l^m_{AB}},T^{AB}).
\end{equation}
In future calculations, it may be necessary to compute derivatives of tensor harmonic fields and evaluate their integrals over $S^2$. For these operations, we refer to Chap. 6 of Ref. \cite{DC} or to Refs. \cite{DB1,CS1}, where the relevant formulas can be found. 

The preceding discussion focused on complex spherical harmonics, nevertheless, employing real spherical harmonics proves simpler for our purposes. The connection between complex and real harmonics is
\begin{equation}
    \begin{aligned}
        &\Bar{Y}_l^o = Y_l^o, \quad \text{for} \quad m=0,\\ 
        &\Bar{Y}_l^m = \frac{(-1)^m}{\sqrt{2}}(Y_l^m + Y_l^{m*}), \quad \text{for} \quad  m>0,\\
        &\Bar{Y}_l^m = \frac{(-1)^m}{i\sqrt{2}}(Y_l^{|m|} - Y_l^{|m|*}), \quad \text{for} \quad m<0.
    \end{aligned}
\end{equation}
All the results commented above also apply to real harmonics. In the subsequent sections and in contrast with Ref. \cite{DC}, we will use real harmonics, omitting from now on the overline notation to simplify the expressions.

When analyzing perturbations about a background where the metric is independent of a specific coordinate, let us say $x$, the expansion in Fourier modes proves highly advantageous. For $x$ living on a compact topology represented by a 1-sphere $S_o^1$, with a period of $L_o$, the $n$ th real Fourier modes, with $n\in\mathbb{N}_0 = \{0,1,2,\cdots\}$, are defined as 
\begin{equation}
    \begin{aligned}
        &Q_0(x) = \frac{1}{\sqrt{L_o}}, \quad \text{for} \quad n=0,\\
        &Q_{n,+}(x) = \sqrt{\frac{2}{L_o}}\cos(\omega_nx), \quad Q_{n,-}(x) = \sqrt{\frac{2}{L_o}}\sin(\omega_nx), \quad \text{for} \quad n\geq1.
    \end{aligned}
\end{equation}
We introduced the positive frequencies $\omega_n = 2\pi n/L_o$. Any arbitrary function $f$ on $S_o^1$ can be decomposed as
\begin{equation}
    f(x) = \sum_{n\in\mathbb{N}_0}\sum_{\lambda} f_{n,\lambda}Q_{n,\lambda}(x), \qquad f_{n,\lambda} = \int_{S_o^1}\text{d}x \;Q_{n,\lambda}(x)f(x).
\end{equation}
We defined $\lambda\in\{+,-\}\simeq\mathbb{Z}_2$ to simplify the notation and used
\begin{equation}
     \int_{S_o^1}\text{d}x \; Q_{n,\lambda}(x)Q_{n',\lambda'}(x) = \delta_{nn'}\delta_{\lambda\lambda'}.
\end{equation}
Note that the Lie derivative of $f$ along the flow generated by $\partial_x$ mixes the $\lambda$ modes of the Fourier expansion and introduces an additional factor of $\lambda\omega_n$, resulting in the expression
\begin{equation}
    \partial_{x}f(x) = \sum_{n\in\mathbb{N}_0}\sum_{\lambda}f_{n,\lambda}\partial_{x}Q_{n,\lambda}(x) = \sum_{n\in\mathbb{N}_0}\sum_{\lambda}\lambda\omega_nf_{n,-\lambda}Q_{n,\lambda}(x).
\end{equation}

\section{Axial Perturbations \label{sec: V}}

Any perturbative quantity in Kantowski-Sachs, such as a symmetric tensor field $T_{ab}$, can be decomposed as 
\begin{equation}
    T_{ab}\text{d}x^a\text{d}x^b = T_{xx}\text{d}x^2 + 2T_{xA}\text{d}x\text{d}x^A + T_{AB}\text{d}x^A\text{d}x^B.
\end{equation}
Clearly, with respect to rotations of $S^2$, $T_{xx}$ represents a scalar, $T_{xA}$ behaves as a covector, and $T_{AB}$ is a symmetric 2-tensor, while all three are scalars on $S_o^1$. As we anticipated, we will treat polar and axial perturbations independently. We center our attention on the axial component of Eq. \eqref{eq: II1}, which can be expressed as,
\begin{equation}
    \label{eq: V1}
    \frac{1}{2}\Delta^2_1[S_0]^{\text{ax}} = \frac{1}{\kappa} \int_{\mathbb{R}}\text{d}t\int_{\sigma_o}\text{d}^3x \left( [h_{ab,t}p^{ab}]^{\text{ax}} - [B^a\Delta[\mathcal{H}_a]]^{\text{ax}} - \frac{N}{2}\Delta^2_1[\mathcal{H}]^{\text{ax}} \right),
\end{equation} 
where the superscript $\text{ax}$ stands for axial part. Our aim is to examine each term of the expression in detail (see Chap. 6 of Ref. \cite{DC} or Ref. \cite{DB3} for a comparative analysis).

\subsection{Hamiltonian formulation \label{ssec: VA}}

We begin the discussion by examining the symplectic form and establishing the perturbative canonical axial variables. An important consideration is that there is no contribution from the perturbative scalar field in the axial case, since it can only be decomposed into polar modes. The axial perturbations of the spatial metric and its momentum are provided by
\begin{equation}
    \label{eq: VA1}
    \begin{aligned}
        &[h_{ab}\text{d}x^a\text{d}x^b]^{\text{ax}} = \sum_{\mathfrak{n}\in\mathfrak{N}_1}\sum_{\lambda} -2h_1^{\mathfrak{n},\lambda}(t) \tensor{X}{_l^m_A}(\theta,\phi)Q_{n,\lambda}(x)\text{d}x\text{d}x^A + \sum_{\mathfrak{n}\in\mathfrak{N}_2}\sum_{\lambda} h_2^{\mathfrak{n},\lambda}(t) \tensor{X}{_l^m_{AB}}(\theta,\phi)Q_{n,\lambda}(x)\text{d}x^A\text{d}x^B,\\
        &\left[\frac{p_{ab}}{\sqrt{\mathfrak{g}}}\text{d}x^a\text{d}x^b\right]^{\text{ax}} = \sum_{\mathfrak{n}\in\mathfrak{N}_1}\sum_{\lambda} -2\Tilde{p}_1^{\mathfrak{n},\lambda}(t) \tensor{X}{_l^m_A}(\theta,\phi)Q_{n,\lambda}(x)\text{d}x\text{d}x^A + \sum_{\mathfrak{n}\in\mathfrak{N}_2}\sum_{\lambda} \Tilde{p}_2^{\mathfrak{n},\lambda}(t) \tensor{X}{_l^m_{AB}}(\theta,\phi)Q_{n,\lambda}(x)\text{d}x^A\text{d}x^B.
    \end{aligned}
\end{equation}
We have defined the set $\mathfrak{N}_k = \{ (n,l,m) \,|\, n\in\mathbb{N}_0,l\in\{k,k+1,k+2,\cdots\},m\in\{-l,\cdots l\} \}$ for $k\geq 1$, and then adopted the abbreviated notation $\mathfrak{n}=(n,l,m)$ for the mode labels. By employing Eq. \eqref{eq: VA1}, the normalization of the basis modes, and a background-dependent and mode-dependent redefinition of the momentum variables, given by
\begin{equation}
    p_1^{\mathfrak{n},\lambda} = \frac{\mathcal{V}}{2\pi L_o}\frac{L_o^2}{p_b^2}l(l+1)\Tilde{p}_1^{\mathfrak{n},\lambda}, \qquad p_2^{\mathfrak{n},\lambda} = \frac{\mathcal{V}}{2\pi L_o}\frac{1}{4p_c^2}\frac{(l+2)!}{(l-2)!}\Tilde{p}_2^{\mathfrak{n},\lambda},
\end{equation}
we attain a concise expression for the symplectic two-form concerning the perturbation. All modes become decoupled and Poisson-commute with the background variables, forming a complete canonical set,
\begin{equation}
    \Theta^{\text{ax}}_{\text{P}} = \sum_{\mathfrak{n}\in\mathfrak{N}_1}\sum_{\lambda} \frac{1}{\kappa}\text{d}h_1^{\mathfrak{n},\lambda}\wedge\text{d}p_1^{\mathfrak{n},\lambda} + \sum_{\mathfrak{n}\in\mathfrak{N}_2}\sum_{\lambda} \frac{1}{\kappa}\text{d}h_2^{\mathfrak{n},\lambda}\wedge\text{d}p_2^{\mathfrak{n},\lambda}, \qquad \{h_i^{\mathfrak{n},\lambda}, p_{i'}^{\mathfrak{n}',\lambda'}\}_{\text{P}} = \kappa \delta_{ii'}\delta_{nn'}\delta_{ll'}\delta_{mm'}\delta_{\lambda\lambda'}, \; \text{for} \; i=1,2.
\end{equation}

Let us consider now the middle term in Eq. \eqref{eq: V1}. It contains two different components, namely, the axial perturbative constraints, which generate perturbative gauge transformations, and their perturbative Lagrange multipliers. They are defined as
\begin{equation}
    \Delta[\mathcal{H}_a]^{\text{ax}} = \left[\Pi^{bc}D_ah_{bc} -2D_c(h_{ab}\Pi^{bc} + g_{ab}p^{bc})\right]^{\text{ax}}, \qquad B_a^{\text{ax}}\text{d}x^a = \sum_{\mathfrak{n}\in\mathfrak{N}_1}\sum_{\lambda} -\kappa h_0^{\mathfrak{n},\lambda}(t)\tensor{X}{_l^m_A}(\theta,\phi)Q_{n,\lambda}(x)\text{d}x^A.
\end{equation}
Upon spatial integration over $\sigma_o$, including the $\kappa$ factor, we obtain the following expression for the constraint term:
\begin{equation}
    \textbf{C}^{\text{ax}}[h_0^{\mathfrak{n},\lambda}] = \sum_{\mathfrak{n}\in\mathfrak{N}_1}\sum_{\lambda}\lambda\omega_n h_0^{\mathfrak{n},-\lambda}\left[ p_1^{\mathfrak{n},\lambda} - 4l(l+1)\frac{L_o^2}{p_b^2}\Omega_bh_1^{\mathfrak{n},\lambda} \right] - \sum_{\mathfrak{n}\in\mathfrak{N}_2}\sum_{\lambda} 2h_0^{\mathfrak{n},\lambda}\left[ p_2^{\mathfrak{n},\lambda} - \frac{(l+2)!}{(l-2)!}(\Omega_b + \Omega_c)\frac{h_2^{\mathfrak{n},\lambda}}{2p_c^2}\right].
\end{equation}

The remaining term in Eq. \eqref{eq: V1} corresponds to the Hamiltonian of the perturbations. The traceless nature of the harmonic basis, which ensures that $\tensor{h}{_a^a}$ and $\tensor{p}{_a^a}$ are zero in the axial case, and the symmetries of the background, simplify its expression to
\begin{equation}
    \begin{aligned}
        &\Delta^2_1[\mathcal{H}]^{\text{ax}} = \left[ \frac{2}{\sqrt{\mathfrak{g}}}p_{ab}p^{ab} -2\sqrt{\mathfrak{g}}G^{ab}\tensor{h}{_a^c}h_{cb} + \frac{\mathcal{H}}{2}h_{ab}h^{ab} + \frac{2}{\sqrt{\mathfrak{g}}}h_{ab}h_{cd}\Pi^{ac}\Pi^{bd} - \frac{2}{\sqrt{\mathfrak{g}}}h^{ab}p_{ab}\tensor{\Pi}{_c^c} + \frac{8}{\sqrt{\mathfrak{g}}}\tensor{\Pi}{_a^c}p_{cb}h^{ab}\right.\\
        &\left.+ 2\sqrt{\mathfrak{g}}\left( D_ah^{ab}D_c\tensor{h}{_b^c} - \frac{3}{4}D_ch_{ab}D^ch^{ab} - h^{ab}D_cD^ch_{ab} + h^{ab}D_bD_c\tensor{h}{_a^c} + h^{ab}D_cD_b\tensor{h}{_a^c} + \frac{1}{2}D_bh_{ac}D^ch^{ab}\right)\right]^{\text{ax}},
    \end{aligned}
\end{equation}
where $G^{xx}=-L_o^2/p_b^2$ is the only nonzero component of the Einstein tensor for the spatial sections $G^{ab}$. Performing the same spatial integration as before, we derive the expression
\begin{equation}
    \begin{aligned}
        \Tilde{H}^{\text{ax}}[\Tilde{N}] &= \sum_{\mathfrak{n}\in\mathfrak{N}_1}\sum_{\lambda}\frac{\Tilde{N}}{\kappa}\left[\frac{p_b^2}{L_o^2}\frac{[p_1^{\mathfrak{n},\lambda}]^2}{l(l+1)} + \left(6\Omega_b^2 + 4\Omega_b\Omega_c + \frac{p_b^2}{L_o^2}l(l+1) + 8\frac{\Tilde{\Pi}^2_{\Phi}}{L_o^2} \right)\frac{L_o^2}{p_b^2}l(l+1)[h_1^{\mathfrak{n},\lambda}]^2 - 4\Omega_b h_1^{\mathfrak{n},\lambda}p_1^{\mathfrak{n},\lambda}\right]\\
        &+ \sum_{\mathfrak{n}\in\mathfrak{N}_2}\sum_{\lambda}\frac{\Tilde{N}}{\kappa}\left[4p_c^2\frac{(l-2)!}{(l+2)!}[p_2^{\mathfrak{n},\lambda}]^2 + \left(2\Omega_b^2 + 4\Omega_c^2 + 4\Omega_b\Omega_c + \omega_n^2p_c^2 + 8\frac{\Tilde{\Pi}_{\Phi}^2}{L_o^2}\right)\frac{1}{4p_c^2}\frac{(l+2)!}{(l-2)!}[h_2^{\mathfrak{n},\lambda}]^2\right.\\ 
        &\left.- 4\Omega_ch_2^{\mathfrak{n},\lambda}p_2^{\mathfrak{n},\lambda} + \lambda\omega_n\frac{(l+2)!}{(l-2)!}h_1^{\mathfrak{n},\lambda}h_2^{\mathfrak{n},-\lambda}\right].
    \end{aligned}
\end{equation}

Therefore, in total, the action of the axial perturbations (at dominant perturbative order) takes the simplified form
\begin{equation}
    \frac{1}{2}\Delta^2_1[S_0]^{\text{ax}} = \int_{\mathbb{R}}\text{d}t\left(\sum_{\mathfrak{n}\in\mathfrak{N}_1}\sum_{\lambda} \frac{1}{\kappa}(h_1^{\mathfrak{n},\lambda})_{,t}p_1^{\mathfrak{n},\lambda} + \sum_{\mathfrak{n}\in\mathfrak{N}_2}\sum_{\lambda} \frac{1}{\kappa}(h_2^{\mathfrak{n},\lambda})_{,t}p_2^{\mathfrak{n},\lambda} - \textbf{C}^{\text{ax}}[h_0^{\mathfrak{n},\lambda}] - \Tilde{H}^{\text{ax}}[\Tilde{N}]\right).
\end{equation}
In future discussions, it will be convenient to split this action into two terms. One will contain the contribution of the modes with $l\geq2$, permitting a unified study of those perturbations, while the other will correspond to the $l=1$ modes, which present certain peculiarities which will be commented later in this work.

\subsection{Perturbative gauge invariant description \label{ssec: VB}} 

To establish a consistent formulation, it is imperative to identify perturbative gauge invariant quantities and express the results in terms of them. As already discussed in Refs. \cite{DC,LMM,DB2}, our approach will involve redefining the phase space variables using perturbative gauge invariants and perturbative constraints (which are also perturbative gauge invariants but constrained to vanish), along with their corresponding conjugate momenta. During the derivation process, the background variables will be considered as fixed, as our primary focus is on addressing the perturbations. Later, once the perturbative description is complete, we will tackle any potential issues of this approach by incorporating a special treatment for the background. Additionally, our analysis will be concentrated on the case $l\geq2$, deferring our comments about the case $l=1$ to the next subsection. Then, at this stage, we can implement a background-dependent and mode-dependent canonical transformation, generated by the following generating function of type $2$: 
\begin{equation}
    \begin{aligned}
        \mathscr{F}_{(2)}[h_i^{\mathfrak{n},\lambda},P_i^{\mathfrak{n},\lambda}] &= \sum_{\mathfrak{n}\in\mathfrak{N}_2}\sum_{\lambda} \left[h_1^{\mathfrak{n},\lambda}P_1^{\mathfrak{n},\lambda} + h_2^{\mathfrak{n},\lambda}P_2^{\mathfrak{n},\lambda} - \frac{\lambda\omega_n}{2}h_2^{\mathfrak{n},\lambda}P_1^{\mathfrak{n},-\lambda} + \frac{1}{4}\frac{(l+2)!}{(l-2)!}\frac{1}{p_c^2}(\Omega_b + \Omega_c)[h_2^{\mathfrak{n},\lambda}]^2\right.\\ 
        &\left.- 2l(l+1)\frac{L_o^2}{p_b^2}\Omega_b\left(\frac{\omega_n^2}{4}[h_2^{\mathfrak{n},\lambda}]^2+\lambda\omega_n h_1^{\mathfrak{n},\lambda}h_2^{\mathfrak{n},-\lambda}\right)\right].
    \end{aligned}
\end{equation}
This transformation mixes $\lambda$ modes, reflecting the behavior of the perturbative constraint. The resulting set satisfies 
\begin{equation}
    \label{eq: VB-1}
    Q_i^{\mathfrak{n},\lambda} = \frac{\partial \mathscr{F}_{(2)}}{\partial P_i^{\mathfrak{n},\lambda}}, \quad p_i^{\mathfrak{n},\lambda} = \frac{\partial \mathscr{F}_{(2)}}{\partial h_i^{\mathfrak{n},\lambda}}, \qquad \{Q_i^{\mathfrak{n},\lambda}, P_{i'}^{\mathfrak{n}',\lambda'}\}_{\text{P}} = \kappa \delta_{ii'}\delta_{nn'}\delta_{ll'}\delta_{mm'}\delta_{\lambda\lambda'} \quad \text{for} \quad i=1,2,
\end{equation}
as it should. In terms of these new variables, the perturbative gauge constraints are simplified to
\begin{equation}
    \label{eq: VB0}
    \textbf{C}^{\text{ax}}[h_0^{\mathfrak{n},\lambda}]\big|_{l\geq2} = - \sum_{\mathfrak{n}\in\mathfrak{N}_2}\sum_{\lambda} 2h_0^{\mathfrak{n},\lambda}P_2^{\mathfrak{n},\lambda},
\end{equation}
whereas the expression for the axial Hamiltonian is given by
\begin{equation}
    \begin{aligned}
        \kappa\textbf{H}^{\text{ax}}\big|_{l\geq2} &= \kappa\Tilde{H}^{\text{ax}}\big|_{l\geq2} + \{\mathscr{F},\Tilde{H}_{\text{KS}}\}_{\text{B}} = \sum_{\mathfrak{n}\in\mathfrak{N}_2}\sum_{\lambda}\left[\frac{p_b^2}{L_o^2}\frac{[P_1^{\mathfrak{n},\lambda}]^2}{l(l+1)} + \omega_n^2p_c^2\frac{(l-2)!}{(l+2)!}\left(P_1^{\mathfrak{n},\lambda} - 4l(l+1)\frac{L_o^2}{p_b^2}\Omega_bQ_1^{\mathfrak{n},\lambda}\right)^2\right.\\
        &+ 4p_c^2\frac{(l-2)!}{(l+2)!}\left(P_2^{\mathfrak{n},\lambda} - \lambda\omega_nP_1^{\mathfrak{n},-\lambda} + 4\lambda\omega_nl(l+1)\frac{L_o^2}{p_b^2}\Omega_bQ_1^{\mathfrak{n},-\lambda} + \frac{1}{p_c^2}\frac{(l+2)!}{(l-2)!}\Omega_bQ_2^{\mathfrak{n},\lambda}\right)P_2^{\mathfrak{n},\lambda}\\
        &\left.+ l(l+1)\frac{L_o^2}{p_b^2}\left(6\Omega_b^2 + 4\Omega_b\Omega_c + 2\frac{p_b^2}{L_o^2} + 8\frac{\Tilde{\Pi}^2_{\Phi}}{L_o^2} + \frac{p_b^2}{L_o^2}(l+2)(l-1)\right)[Q_1^{\mathfrak{n},\lambda}]^2 - 4\Omega_b Q_1^{\mathfrak{n},\lambda}P_1^{\mathfrak{n},\lambda}\right].
    \end{aligned}
\end{equation}
 In this derivation, we have taken into account the background dependence of the canonical transformation, making it necessary to introduce a correction term in the Hamiltonian owing to the variation of the generating function produced by the background\footnote{This term reproduces the time derivative of the generating function when we consider the background as a time-dependent quantity.}. We have decided to omit the type of generating function when correcting the Hamiltonian since the distinction is irrelevant for our discussion, inasmuch as different types differ only by products of perturbative variables, which can be ignored in taking Poisson brackets with the background. By redefining the Lagrange multipliers \cite{LMM}, we can eliminate nonrelevant terms from the Hamiltonian. This redefinition holds valid within our perturbative truncation order in the action, and results in
\begin{equation}
    \label{eq: VB1}
    \begin{aligned}
        \kappa\Tilde{\textbf{H}}^{\text{ax}}\big|_{l\geq2} = \sum_{\mathfrak{n}\in\mathfrak{N}_2}\sum_{\lambda}&\left[\frac{p_b^2}{L_o^2}\frac{[P_1^{\mathfrak{n},\lambda}]^2}{l(l+1)} + l(l+1)\frac{L_o^2}{p_b^2}\left(8\Omega_b^2 + 8\Omega_b\Omega_c + 4\frac{p_b^2}{L_o^2} + \frac{p_b^2}{L_o^2}(l+2)(l-1)\right)[Q_1^{\mathfrak{n},\lambda}]^2\right.\\
        &\left.+ \omega_n^2p_c^2\frac{(l-2)!}{(l+2)!}\left(P_1^{\mathfrak{n},\lambda} - 4l(l+1)\frac{L_o^2}{p_b^2}\Omega_bQ_1^{\mathfrak{n},\lambda}\right)^2 - 4\Omega_b Q_1^{\mathfrak{n},\lambda}P_1^{\mathfrak{n},\lambda}\right].
    \end{aligned}
\end{equation}
This Hamiltonian only depends on the perturbative gauge invariant pairs $(Q_1^{\mathfrak{n},\lambda}, P_1^{\mathfrak{n},\lambda})$ and the geometric background variables\footnote{Using Eqs. \eqref{eq: VB-1} and \eqref{eq: VB0}, it is straightforward to see that the variables  $Q_1^{\mathfrak{n},\lambda}$ and $P_1^{\mathfrak{n},\lambda}$ commute under Poisson brackets with the generator of the perturbative gauge transformations, namely the perturbative constraint $\textbf{C}^{\text{ax}}[h_0^{\mathfrak{n},\lambda}]\big|_{l\geq2}$. Consequently, they are perturbative gauge invariants.}. On the other hand, the redefined Lagrange multipliers are
\begin{equation}
    \begin{aligned}
        &\Tilde{\textbf{N}} = \Tilde{N}\left[1 + \frac{\epsilon^2}{2\kappa L_o}\sum_{\mathfrak{n}\in\mathfrak{N}_2}\sum_{\lambda}l(l+1)\frac{p_b^2}{L_o^2}[Q_1^{\mathfrak{n},\lambda}]^2\right],\\
        &\textbf{h}_0^{\mathfrak{n},\lambda} = h_0^{\mathfrak{n},\lambda} - \frac{\Tilde{N}}{\kappa}2p_c^2\frac{(l-2)!}{(l+2)!}\left[P_2^{\mathfrak{n},\lambda} - \lambda\omega_nP_1^{\mathfrak{n},-\lambda} + 4\lambda\omega_nl(l+1)\frac{L_o^2}{p_b^2}\Omega_bQ_1^{\mathfrak{n},-\lambda} + \frac{1}{p_c^2}\frac{(l+2)!}{(l-2)!}\Omega_bQ_2^{\mathfrak{n},\lambda}\right].
    \end{aligned}
\end{equation}
We highlight the appearance of the factor $\epsilon^2$ in the modification of the lapse function, which clearly shows that this change in lapse is of quadratic perturbative order \cite{LMM}.

We now focus our discussion on refining the expression of the Hamiltonian obtained in Eq. \eqref{eq: VB1}. To achieve this goal, we introduce a canonical transformation defined by the following generating function of type $3$:
\begin{equation}
    \mathscr{G}_{(3)}[\Tilde{Q}_i^{\mathfrak{n},\lambda},P_i^{\mathfrak{n},\lambda}] = -\sum_{\mathfrak{n}\in\mathfrak{N}_2}\sum_{\lambda}\left[ \sqrt{\frac{(l-2)!}{(l+2)!}\left[(l+2)(l-1)\frac{p_b^2}{L_o^2} + \omega_n^2p_c^2\right]}\Tilde{Q}_1^{\mathfrak{n},\lambda}P_1^{\mathfrak{n},\lambda} + \Tilde{Q}_2^{\mathfrak{n},\lambda}P_2^{\mathfrak{n},\lambda} + \frac{F_{nl}}{2}[\Tilde{Q}_1^{\mathfrak{n},\lambda}]^2 \right].
\end{equation}
To change from the set $\{Q_i^{\mathfrak{n},\lambda},P_i^{\mathfrak{n},\lambda}\}_{i=1}^2$ to the set $\{\Tilde{Q}_i^{\mathfrak{n},\lambda},\Tilde{P}_i^{\mathfrak{n},\lambda}\}_{i=1}^2$, we use the definitions
\begin{equation}
     Q_i^{\mathfrak{n},\lambda} = -\frac{\partial \mathscr{G}_{(3)}}{\partial P_i^{\mathfrak{n},\lambda}}, \quad \Tilde{P}_i^{\mathfrak{n},\lambda} = -\frac{\partial \mathscr{G}_{(3)}}{\partial \Tilde{Q}_i^{\mathfrak{n},\lambda}}, \qquad F_{nl} = \frac{(l+2)(l-1)p_b^2(\Omega_b - \Omega_c)}{(l+2)(l-1)p_b^2 +  \omega_n^2p_c^2 L_o^2} - 4\Omega_b\left[1+\frac{\omega_n^2p_c^2L_o^2}{(l+2)(l-1)p_b^2}\right].
\end{equation}
Considering the variation of the above generating function coming from its background dependence (and disregarding again the distinction between different types of generating functions), the axial Hamiltonian becomes
\begin{equation}
    \begin{aligned}
        \kappa\mathscr{H}^{\text{ax}}\big|_{l\geq2} &= \kappa\Tilde{\textbf{H}}^{\text{ax}}\big|_{l\geq2} + \{\mathscr{G},\Tilde{H}_{\text{KS}}\}_{\text{B}} = \sum_{\mathfrak{n}\in\mathfrak{N}_2}\sum_{\lambda}\left[\left((l+2)(l-1)\frac{p_b^2}{L_o^2} + \omega_n^2p_c^2 + \frac{(l+2)(l-1)p_b^4}{(l+2)(l-1)p_b^2L_o^2 + \omega_n^2p_c^2 L_o^4}\right.\right.\\
        &\left.\left.- \frac{[(l+2)(l-1)p_b^2]^2-2\omega_n^2(l+2)(l-1)p_c^2p_b^2 L_o^2}{[(l+2)(l-1)p_b^2 + \omega_n^2p_c^2 L_o^2]^2}(\Omega_b-\Omega_c)^2\right)[\Tilde{Q}_1^{\mathfrak{n},\lambda}]^2 + [\Tilde{P}_1^{\mathfrak{n},\lambda}]^2\right].
    \end{aligned}
\end{equation}
This expression is significantly more manageable than Eq. \eqref{eq: VB1}, featuring only diagonal quadratic contributions and gathering all the background dependence on a single factor. Moreover, it exhibits a notable resemblance to the one in Ref. \cite{AT}, which prompts us to rewrite the Hamiltonian in terms of the quantities
\begin{equation}
    k^2 = (l+2)(l-1) + \omega_n^2, \qquad \hat{l} = \frac{\sqrt{(l+2)(l-1)}}{k}, \qquad k^2b_{\hat{l}}^2 = (l+2)(l-1)\frac{p_b^2}{L_o^2} + \omega_n^2p_c^2.
\end{equation}
We also consider the same transformation discussed in Ref. \cite{AT}, generated by a generating function of type $2$, namely 
\begin{equation}
    \label{eq: VB2}
     \mathscr{K}_{(2)}[\Tilde{Q}_i^{\mathfrak{n},\lambda},\mathscr{P}_i^{\mathfrak{n},\lambda}] = \sum_{\mathfrak{n}\in\mathfrak{N}_2}\sum_{\lambda}
    \left[\sqrt{b_{\hat{l}}}\Tilde{Q}_1^{\mathfrak{n},\lambda}\mathscr{P}_1^{\mathfrak{n},\lambda} + \Tilde{Q}_2^{\mathfrak{n},\lambda}\mathscr{P}_2^{\mathfrak{n},\lambda} - \frac{b_{\hat{l}}'}{4b_{\hat{l}}}[\Tilde{Q}_1^{\mathfrak{n},\lambda}]^2\right],
\end{equation}
with $b_{\hat{l}}' = \{b_{\hat{l}},\Tilde{H}_{\text{KS}}\}_{\text{B}}$ and $\{\mathscr{Q}_i^{\mathfrak{n},\lambda},\mathscr{P}_i^{\mathfrak{n},\lambda}\}_{i=1}^2$ the new set of canonical variables, determined by
\begin{equation}
\mathscr{Q}_i^{\mathfrak{n},\lambda} = \frac{\partial \mathscr{K}_{(2)}}{\partial \mathscr{P}_i^{\mathfrak{n},\lambda}}, \quad \Tilde{P}_i^{\mathfrak{n},\lambda} = \frac{\partial \mathscr{K}_{(2)}}{\partial \Tilde{Q}_i^{\mathfrak{n},\lambda}}.
\end{equation}
We then obtain (with the same remarks about the type of generating function as above)
\begin{equation}
    \label{eq: VB3}
    \kappa\Tilde{\mathscr{H}}^{\text{ax}}\big|_{l\geq2} = \kappa\mathscr{H}^{\text{ax}}\big|_{l\geq2} + \{\mathscr{K},\Tilde{H}_{\text{KS}}\}_{\text{B}} = \sum_{\mathfrak{n}\in\mathfrak{N}_2}\sum_{\lambda}b_{\hat{l}}\left([\mathscr{P}_1^{\mathfrak{n},\lambda}]^2 + [k^2 + s_{\hat{l}}][\mathscr{Q}_1^{\mathfrak{n},\lambda}]^2\right),
\end{equation}
where
\begin{equation}\label{ese}
    s_{\hat{l}} = s_{\hat{l}}^{\text{A}} + 2\frac{\hat{l}^2}{b_{\hat{l}}^4}\frac{p_b^2}{L_o^2}\left[(\Omega_b-\Omega_c)^2 + \frac{1}{2}\frac{p_b^2}{L_o^2}\right] - 3\frac{\hat{l}^4}{b_{\hat{l}}^6}\frac{p_b^4}{L_o^4}(\Omega_b-\Omega_c)^2.
\end{equation}
The first contribution corresponds to the mass term of the modes of a free Klein-Gordon field in a Kantowski-Sachs geometry. Its expression can be computed using the commutation relationships of the background, as
\begin{equation}
    s_{\hat{l}}^{\text{A}} = \frac{4}{b_{\hat{l}}^2}\left[\frac{p_b^2}{L_o^2} + \Omega_b^2\right] - 2\frac{\hat{l}^2}{b_{\hat{l}}^4}\frac{p_b^2}{L_o^2}\left[3(\Omega_b-\Omega_c)^2 + \frac{p_b^2}{L_o^2} + (\Omega_b^2-\Omega_c^2)\right] + 5\frac{\hat{l}^4}{b_{\hat{l}}^6}\frac{p_b^4}{L_o^4}(\Omega_b-\Omega_c)^2.
\end{equation}
The two additional terms in Eq. \eqref{ese} are interpreted as corrections that arise because the Kantowski-Sachs perturbative modes do not have the same nature as the modes of a free Klein-Gordon field. The last canonical transformation that we have performed is crucial for the remaining part of our work, as we will see. Finally, as the discussed axial, perturbative gauge invariant variables are concerned, the action at the first relevant perturbative order is given by
\begin{equation}
    \frac{1}{2}\Delta^2_1[S_0]^{\text{ax}}\big|_{l\geq2} = \int_{\mathbb{R}}\text{d}t\left(\sum_{\mathfrak{n}\in\mathfrak{N}_2}\sum_{\lambda}\frac{1}{\kappa}(\mathscr{Q}_1^{\mathfrak{n},\lambda})_{,t}\mathscr{P}_1^{\mathfrak{n},\lambda} + \sum_{\mathfrak{n}\in\mathfrak{N}_2}\sum_{\lambda}\frac{1}{\kappa}(\mathscr{Q}_2^{\mathfrak{n},\lambda})_{,t}\mathscr{P}_2^{\mathfrak{n},\lambda} - \textbf{C}^{\text{ax}}[\textbf{h}_0^{\mathfrak{n},\lambda}]\big|_{l\geq2} - \Tilde{\mathscr{H}}^{\text{ax}}[\Tilde{\textbf{N}}]\big|_{l\geq2}\right).
\end{equation}

\subsection{Modes with l=1 \label{ssec: VC}}

For the remaining perturbative modes with $l=1$, which we did not consider in the previous subsection, a unique set of canonical variables is provided by $\{h_1^{\mathfrak{n},\lambda}, p_1^{\mathfrak{n},\lambda}\}_{l=1}$. Employing the same rationale as before, we introduce the following canonical transformation\footnote{It is understood that, in this subsection, all expressions refer to the case $l=1$. To maintain a clear notation, we will omit the evaluation at this value of $l$ in the perturbative modes and use the same notation as before.}:
\begin{equation}
    Q_1^{\mathfrak{n},\lambda} = \frac{\partial \Tilde{\mathscr{F}}_{(2)}}{\partial P_1^{\mathfrak{n},\lambda}}, \quad p_1^{\mathfrak{n},\lambda} = \frac{\partial \Tilde{\mathscr{F}}_{(2)}}{\partial h_1^{\mathfrak{n},\lambda}}, \qquad \Tilde{\mathscr{F}}_{(2)}[h_1^{\mathfrak{n},\lambda},P_1^{\mathfrak{n},\lambda}] = \sum_{\mathfrak{n}; l=1}\sum_{\lambda} \left[h_1^{\mathfrak{n},\lambda}P_1^{\mathfrak{n},\lambda} + 4\frac{L_o^2}{p_b^2}\Omega_b[h_1^{\mathfrak{n},\lambda}]^2\right].
\end{equation}
The perturbative constraints and the contribution to the axial Hamiltonian, expressed in the set $\{Q_1^{\mathfrak{n},\lambda}, P_1^{\mathfrak{n},\lambda}\}_{l=1}$, are 
\begin{equation}
    \begin{aligned}
        &\textbf{C}^{\text{ax}}[h_0^{\mathfrak{n},\lambda}]\big|_{l=1} = -\sum_{\mathfrak{n};l=1}\sum_{\lambda}\lambda\omega_n h_0^{\mathfrak{n},\lambda}P_1^{\mathfrak{n},-\lambda},\\
        &\kappa\textbf{H}^{\text{ax}}\big|_{l=1} = \kappa\Tilde{H}^{\text{ax}}\big|_{l=1} + \{\Tilde{\mathscr{F}},\Tilde{H}_{\text{KS}}\}_{\text{B}} = \sum_{\mathfrak{n};l=1}\sum_{\lambda}\left[\frac{1}{2}\frac{p_b^2}{L_o^2}[P_1^{\mathfrak{n},\lambda}]^2 + 4\Omega_b Q_1^{\mathfrak{n},\lambda}P_1^{\mathfrak{n},\lambda}\right].
    \end{aligned}
\end{equation}
Based on these expressions, we can divide our analysis into two different cases. If $n$ is nonzero, we can conveniently redefine the perturbative Lagrange multipliers in a manner similar to that performed in our discussion above, namely
\begin{equation}
    \textbf{h}_0^{\mathfrak{n},\lambda} = h_0^{\mathfrak{n},\lambda} - \frac{\Tilde{N}}{\lambda\omega_n\kappa}\left[\frac{1}{2}\frac{p_b^2}{L_o^2}P_1^{\mathfrak{n},-\lambda} + 4\Omega_b Q_1^{\mathfrak{n},-\lambda}\right].
\end{equation}
Ultimately, the $n\neq0$ modes have no physical relevance, being purely gauge, as evidenced by their contribution to the action,
\begin{equation}
    \frac{1}{2}\Delta^2_1[S_0]^{\text{ax}}\big|_{l=1}^{n\neq0} = \int_{\mathbb{R}}\text{d}t\left.\left(\sum_{\mathfrak{n};l=1}\sum_{\lambda}\frac{1}{\kappa}(Q_1^{\mathfrak{n},\lambda})_{,t}P_1^{\mathfrak{n},\lambda} - \textbf{C}^{\text{ax}}[\textbf{h}_0^{\mathfrak{n},\lambda}]\big|_{l=1}\right)\right|_{n\neq0}.
\end{equation}

When $n$ is equal to zero, on the other hand, three distinct modes appear, corresponding to $m\in\{-1,0,1\}$, and no constraint is imposed on them. In this situation, we can carry out another canonical transformation, generated by
\begin{equation}
\Tilde{\mathscr{G}}_{(3)}[\Tilde{Q}_1^{\mathfrak{n},\lambda},P_1^{\mathfrak{n},\lambda}] = -\sum_{\mathfrak{n};l=1}\sum_{\lambda}\left.\left[\frac{p_b}{L_o}\Tilde{Q}_1^{\mathfrak{n}\lambda}P_1^{\mathfrak{n},\lambda} + (\Omega_b - \Omega_c)[\Tilde{Q}_1^{\mathfrak{n}\lambda}]^2\right]\right|_{n=0},
\end{equation}
with
\begin{equation}
    Q_1^{\mathfrak{n},\lambda} = -\frac{\partial \Tilde{\mathscr{G}}_{(3)}}{\partial P_1^{\mathfrak{n},\lambda}}, \quad \Tilde{P}_1^{\mathfrak{n},\lambda} = -\frac{\partial \Tilde{\mathscr{G}}_{(3)}}{\partial \Tilde{Q}_1^{\mathfrak{n},\lambda}}.
\end{equation}
The corresponding Hamiltonian, expressed in terms of the new set of variables, $\{\Tilde{Q}_1^{\mathfrak{n},\lambda},\Tilde{P}_1^{\mathfrak{n},\lambda}\}_{l=1}^{n=0}$, is
\begin{equation}
    \kappa\Tilde{\textbf{H}}^{\text{ax}}\big|_{l=1}^{n=0} = \kappa\textbf{H}^{\text{ax}}\big|_{l=1}^{n=0} + \{\Tilde{\mathscr{G}},\Tilde{H}_{\text{KS}}\}_{\text{B}} = \sum_{\mathfrak{n};l=1}\sum_{\lambda}\left.\left[\frac{1}{2}[\Tilde{P}_1^{\mathfrak{n},\lambda}]^2 + 2\left(\frac{p_b^2}{L_o^2} - (\Omega_b-\Omega_c)^2\right)[\Tilde{Q}_1^{\mathfrak{n},\lambda}]^2\right]\right|_{n=0}.
\end{equation}
This formula can be interpreted as the Hamiltonian of a system of three generalized harmonic oscillators with background-dependent frequencies. If we incorporate the background contribution to the total Hamiltonian constraint, it is not difficult to prove that, up to our perturbative order, the resulting squared frequencies can be negative if the zero mode for the scalar field vanishes. In such a scenario, a nonconventional quantization approach should be adopted to mitigate issues related to these negative squared frequencies, such as tachyonic behavior. Nevertheless, their quantization will not affect the ultraviolet behavior of the perturbations, since they represent a finite number of degrees of freedom. In the following, we ignore these modes and assume that they are taken to vanish.

\section{Hybrid quantization in Loop Quantum Cosmology \label{sec: VI}}

After completing the Hamiltonian description of the perturbations, we proceed to the quantization of our perturbed Kantowski-Sachs spacetime. For this aim, the main issue not yet addressed is the recovery of a global canonical structure, encompassing both background and perturbation variables. In Appendix \ref{sec: app}, we extensively discuss this matter. The result is that the only implication of the passage to a canonical set for the combined perturbed system is a change in our background variables, that acquire suitable quadratic corrections in the perturbations. For practical purposes, this change reduces to a change of notation for the background variables, which will now appear with an overhead bar. Consequently, in the following, any quantity defined on the background, when adorned with an overhead bar, refers to that expression in terms of the newly corrected variables. 

To achieve a quantization that includes both perturbative and nonperturbative degrees of freedom, we adopt the LQC-Hybrid procedure, a well-established technique in the LQG community. This choice is motivated by the success of the approach in the treatment of other cosmological scenarios of interest, such as Bianchi I or standard cosmology \cite{hyb-review}. A key hypothesis of the hybrid method is the assumption of different quantization approaches for perturbative and nonperturbative variables. Considering that the most relevant quantum geometry aspects impact the background spacetime, a specialized quantization is necessary. Meanwhile, the quantum effects of the perturbations can be addressed using a more standard approach. Throughout this work, we will adopt the LQC and Fock (or Schrödinger) representations for the background and the perturbations, respectively. It is important to note that the hybrid method can be applied to any other type of quantum approaches. Despite being quantized differently, the perturbative and nonperturbative sectors collectively form a symplectic manifold, which is quantized as a whole. Subsequently, the method involves imposing classical constraints quantum mechanically, following Dirac's proposals, and the viability of their implementation is guaranteed thanks to the treatment that we have given to them in the previous sections. These constraints are represented as operators that annihilate physical states in the quantum theory, affecting both sectors simultaneously. Notably, they encode backreaction of the perturbations on the background, provided our analysis remains within the quadratic perturbative truncation of the action that we are considering. In this section, we will examine the details of each of the aforementioned quantum representations.

\subsection{LQC representation of the background \label{ssec: VIA}}

The fundamental variables for LQG consist of holonomies of a $SU(2)$ connection and fluxes of the densitized triad \cite{A&L,Thiemann}. In Kantowski-Sachs, relevant holonomies are expressed in terms of complex exponentials of the form $\mathcal{N}_{\mu_j} = e^{i\bar{j}\mu_j/2}$, where $\mu_j \in \mathbb{R}$ represents a coordinate length parameter, $\bar{j}=\bar{b}$ or $\bar{c}$, and fluxes over surfaces are defined in terms of the variables $\bar{p}_j$ \cite{GBA}. When examining the holonomy-flux algebra, it becomes evident that its representation is discrete, leading to a limitation in defining connection variables as operators in the quantum theory. Given that our Hamiltonian depends on them, this poses a challenge. However, this problem can be resolved through an established regularization procedure, which can be summarized in the following recipe: $\bar{j} \rightarrow \sin(\delta_j \bar{j})/\delta_j$, where $\delta_j$ are two regularization parameters of quantum origin. By employing an extended formulation in which this regularization parameters, along with their corresponding conjugate momenta $p_{\delta_j}$, are considered as canonical variables, we can derive effective results for the dynamical trajectories within this extended phase space. In this scenario, the dynamics unfold according to the effective extended Hamiltonian \cite{GBA}:
\begin{equation}
    \label{eq: VIA1}
    \Tilde{H}^{\text{eff}}_{\text{ext}}[\tilde{N},\lambda_b,\lambda_c] = \tilde{H}_{\text{KS}}^{\text{reg}}[\tilde{N}] + \Psi_b[\lambda_b] + \Psi_c[\lambda_c].
\end{equation}
In the above expression, the first term represents the regularized version of Eq. \eqref{eq: III1}, while each of the two additional terms involves a Lagrange multiplier $\lambda_j$ and a constraint $\Psi_j$. These two extra constraints enforce a relationship between the regularization parameters and the original phase space variables, dependent on the prescription under consideration. Among the proposals found in the literature, one notable suggestion is put forth by Ashtekar, Olmedo, and Singh (AOS) \cite{AOS}. In their proposal, these authors suggest fixing the values of the regularization parameters to a certain function of $\bar{\Omega}_c$ (a conserved quantity along physical trajectories that classically can be identified with the ADM mass), at least asymptotically for large values of this quantity. The strength of this prescription lies in the good physical properties found when the spacetime is compared with the interior of a black hole \cite{AOS, AOS2}, displaying in particular small quantum corrections near the black hole horizon. Kantowski-Sachs results in GR are recovered when the regularization parameters approach zero, and a suitable partial gauge-fixing is enforced by the conditions $\lambda_j = 0$. A more detailed discussion of the extended formulation can be found in Ref. \cite{BH_GAB}. 

Under these considerations, an extended kinematic representation of the model can be formulated by applying loops techniques to quantize the geometric degrees of freedom, while a continuous Schrödinger representation is carried out for the quantization of the homogeneous scalar field and the $\delta$-parameters. For the geometry, the Hilbert space, denoted as $\mathcal{H}_{\text{LQC}}^{\text{kin}}$, is constructed in the densitized triad representation by completing each copy of the holonomy-flux algebra with respect to the discrete product and taking the tensor product of the two individual Hilbert spaces. For a more straightforward description, it proves beneficial to redefine our basis in terms of the eigenvalues of the rescaled quantities $\tilde{p}_j = \bar{p}_j/\delta_j$. Explicitly, for the holonomy elements and flux operators we have  
\begin{equation}
\hat{\mathcal{N}}_{\delta_b} |\tilde{\mu}_b \rangle = |\tilde{\mu}_b + 1 \rangle, \qquad \hat{\mathcal{N}}_{\delta_c} |\tilde{\mu}_c \rangle = |\tilde{\mu}_c + 1 \rangle, \qquad \hat{\Tilde{p}}_b |\tilde{\mu}_b\rangle = \frac{\gamma \tilde{\mu}_b}{2} |\tilde{\mu}_b\rangle, \qquad \hat{\Tilde{p}}_c |\tilde{\mu}_c\rangle = \gamma \tilde{\mu}_c  |\tilde{\mu}_c\rangle. 
\end{equation} 
Hence, we observe that the action of $\hat{\mathcal{N}}_{\delta_j}$ and $\hat{\tilde{p}}_j$ is independent of the $\delta$-parameters of the model. Since the representation of the remaining canonical pairs does not pose any problem, we proceed to present the resulting kinematic Hilbert space as the tensor product $\mathcal{H}_{\text{ext}}^{\text{kin}}$ = $\mathcal{H}_{\text{LQC}}^{\text{kin}} \otimes_j L^2(\mathbb{R},\text{d}\delta_j) \otimes L^2(\mathbb{R},\text{d}\Phi)$, for which a convenient basis is provided by the states $|\Tilde{\mu}_b,\Tilde{\mu}_c,\delta_b,\delta_c,\Phi\rangle$ obtained from the tensor product of the respective individual bases. Here, the labels $\Tilde{\mu}_j$ are normalized to the Kronecker delta, while the rest are normalized to the Dirac delta. 

Only the first term of Eq. \eqref{eq: VIA1} requires a detailed quantum treatment. The implementation of the other two constraints at the quantum level is straightforward. Thanks to the regularization process that we have undertaken, we can derive a well-defined quantum expression for $\tilde{H}_{\text{KS}}^{\text{reg}}[\tilde{N}]$, upon multiplication by the lapse function, which is represented as
\begin{equation}
    \hat{H}_{\text{KS}} = -L_o\left[\hat{h}_{\text{KS}} - \frac{4}{L_o^2}\hat{\Pi}_{\Phi}^2\right], \qquad \hat{h}_{\text{KS}} = \hat{\Omega}_b^2 + \hat{\delta}_b^2\frac{\hat{\Tilde{p}}_b^2}{L_o^2} + 2\hat{\Omega}_b\hat{\Omega}_c, \qquad \hat{\Pi}_{\Phi} = -i\partial_{\Phi}.
\end{equation}
Regarding the $\hat{h}_{\text{KS}}$ operator, its essential self-adjointness was demonstrated in Ref. \cite{GBA}, and it is primarily composed of $\hat{\Omega}_j$ operators, which classically represent the regularized version of $\bar{\Omega}_j$. They are defined as 
\begin{equation}
    \hat{\Omega}_j = \frac{1}{2\gamma L_o}|\hat{\tilde{p}}_j|^{1/2}\left[\widehat{\sin(\delta_j \bar{j})}\widehat{\text{sign}(\tilde{p}_j)}+\widehat{\text{sign}(\tilde{p}_j)}\widehat{\sin(\delta_j \bar{j})}\right]|\hat{\tilde{p}}_j|^{1/2}, \qquad \widehat{\sin{(\delta_ j \bar{j})}} = \frac{1}{2i} (\hat{\mathcal{N}}_{2\delta_j}-\hat{\mathcal{N}}_{-2\delta_j}), 
\end{equation}
where we have employed the Martin-Benito-Mena Marugan-Olmedo (MMO) symmetrization prescription \cite{MMO}, a method that has proven successful in LQC for various cosmological scenarios. An important feature of these operators, inherent from the MMO factor ordering prescription, is that their action leaves invariant the Hilbert subspaces $\prescript{(2)}{}{\mathcal{H}}_{\epsilon_j}^{\pm}$ formed by states with support on semilattices $\prescript{(2)}{}{\mathcal{L}}_{\epsilon_j}^{\pm} = \{\pm(\epsilon_j + 2n)\,|\,n\in\mathbb{N}_0\}$, where $\epsilon_j \in (0,2]$. As a consequence, instead of working with $\mathcal{H}_{\text{LQC}}^{\text{kin}}$, this property allows us to restrict our study to one of these separable superselected Hilbert subspaces. Remarkably, we note that $\hat{\tilde{p}}_j$ is a positive (negative) operator in $\prescript{(2)}{}{\mathcal{H}}_{\epsilon_j}^{+}$ ($\prescript{(2)}{}{\mathcal{H}}_{\epsilon_j}^{-}$). Therefore, on these superselected subspaces, we can always represent any fractional power of the absolute value of $\bar{p}_j$, including negative powers, without any obstruction. For simplicity, and without loss of generality, we choose to work with $\prescript{(2)}{}{\mathcal{H}}_{\epsilon_b}^{+}\otimes\prescript{(2)}{}{\mathcal{H}}_{\epsilon_c}^{+}$ in the following. 

Additionally, in Ref. \cite{GBA} a method for computing the states annihilated under the action of $\hat{h}_{\text{KS}}$ was presented (up to a multiplicative factor of $L_o^2$, which is not relevant for our discussion). The habitat of the mentioned states is the algebraic dual of the eigenstates of $\hat{\tilde{p}}_j$, for each generalized eigenspace of the operators $\hat{\delta}_j$. Under these considerations, solving the Hamiltonian constraint operator is not a complicated task. To proceed, for instance, the constrained system can be deparametrized by choosing $\Phi$ as an internal time. This procedure yields two sectors corresponding to positive and negative frequencies, establishing a notion of dynamical evolution, $\mathbb{R}\ni\Phi\mapsto|\Xi_{\Phi}\rangle=|\Xi,\Phi\rangle$, with $|\Xi\rangle\in\mathcal{H}_{\text{LQC}}^{\text{kin}}\otimes_jL^2(\mathbb{R},\text{d}\delta_j)$, where time translations manifest as unitary transformations given by
\begin{equation}
    |\Xi_{\Phi_o}\rangle \mapsto |\Xi_{\Phi}\rangle = e^{\pm i L_o \sqrt{\hat{h}_{\text{KS}}}(\Phi - \Phi_o)/2}\,|\Xi_{\Phi_o}\rangle,
\end{equation}
with $\sqrt{\hat{h}_{\text{KS}}}$ (or, more rigorously, the square root of the positive part of $\hat{h}_{\text{KS}}$) providing the generator of time transformations \cite{Prescrip_G}. The imposition of the two remaining quantum constraints ensures that the dependence of the state on $\delta_j$ is only through the AOS proposal or, strictly speaking, through the proposal incorporated in $\Psi_j$. In this sense, we notice that the extended formalism can be easily applied to other scenarios. It suffices to introduce a Dirac delta for each regularization parameter to enforce the desired relationship in the wave function $\Xi_{\Phi_o} = \Xi_{\Phi_o}(\Tilde{\mu}_b,\Tilde{\mu}_c,\delta_b,\delta_c)$, associated with the state $|\Xi_{\Phi_o}\rangle$, similar to what was done in Refs. \cite{BH_GAB, GBA}.
For consistency, we only need that his relationship is established with Dirac observables of the quantum theory, namely, the counterpart of classical constants of motion.

\subsection{Representation of the perturbations and quantum constraints \label{ssec: VIB}}

In the LQC-Hybrid approach, perturbative modes can be treated using a Fock quantization. This choice is motivated by the recovery of a quantum field theory in the curved background geometry \cite{hyb-review}. We adopt this type of representation for the perturbative gauge invariant variables. Their Fock representation turns out to be unique up to unitary equivalence if one demands certain reasonable requirements, namely, the invariance of the vacuum under spatial isometries, and the unitary implementation of the dynamics of the perturbations when the background is regarded as classical \cite{AT}. It is worth noting that, from a rigorous mathematical point of view, the proof of this result uses that the zero modes can be isolated because the spatial sections are compact, a property that reaffirms our selection of $\sigma_o$. The demonstration of this uniqueness statement is straightforward when one compares the Hamiltonian in Eq. \eqref{eq: VB3} with the system studied in Ref. \cite{AT}. We stress the importance of the canonical transformation described in Eq. \eqref{eq: VB2}, as it is pivotal in achieving a unique quantization with unitary dynamics. Consequently, the annihilation and creation variables defined by the relations
\begin{eqnarray}
    &a_{\mathfrak{n},\lambda} = f_{nl}(t)\mathscr{Q}_1^{\mathfrak{n},\lambda} + g_{nl}(t)\mathscr{P}_1^{\mathfrak{n},\lambda}, \qquad a_{\mathfrak{n},\lambda}^* = f_{nl}^*(t)\mathscr{Q}_1^{\mathfrak{n},\lambda} + g_{nl}^*(t)\mathscr{P}_1^{\mathfrak{n},\lambda}, \\
&f_{nl}(t)g_{nl}^*(t) - f_{nl}^*(t)g_{nl}(t) = -i,
\end{eqnarray}
are suitable to characterize the set of invariant Fock representations. Note that these variables do not mix modes and that the functions of time $f_{nl}(t)$ and $g_{nl}(t)$ that parametrize the variables may depend only on the mode labels $n$ and $l$, but not on $m$. In addition, these functions have to satisfy some conditions in order to implement the dynamical evolution of the creation and annihilation variables as unitary quantum transformations, an issue for which we refer to the analysis carried out in Ref. \cite{AT}.

Let us call $\mathcal{F}^{\text{ax}}_P$ the Fock space for a given set of creation and annihilation operators belonging to the above unitarily equivalent family of representations for the perturbative modes. Then, a basis of states is provided by the occupancy-number state $|\mathcal{N}\rangle$, where $\mathcal{N}$ denotes an array of occupancy-numbers for the modes with a finite number of nonzero entries. Once the Fock space for the axial perturbations is defined, the implementation of the Hamiltonian and the perturbative constraints, considered as operators, is not very difficult to carry out. On one hand, the quantum implementation of the axial constraints is immediate. Composed by products of momentum operators multiplied by their corresponding Lagrange multipliers, their action can be described by a set of generalized derivatives. This means that the states annihilated by them do not depend on the conjugate variables, which, in the classical theory, were indeed purely gauge quantities. We emphasize that these restrictions on the quantum states come from the quantum theory and not as a reduction of the classical theory. 

On the other hand, the quantum implementation of the axial Hamiltonian is a task that requires more effort. Given that the Hamiltonian is quadratic in the perturbative gauge invariants, for which we adopt a Fock representation, the only real challenge in representing it as a quantum operator lies in implementing its background-dependent coefficients. Using the results presented in the previous section regarding the background quantization, we observe that the two types of background factors that appear in the axial Hamiltonian \eqref{eq: VB3}, namely $\bar{b}_{\hat{l}}$ and $\bar{s}_{\hat{l}}$, depend on the background variables just through $\bar{p}_j$ and/or $\bar{\Omega}_j$. These quantities are perfectly representable as essentially self-adjoint operators acting on the Hilbert space, and the operator for $\bar{p}_j$ results to be positive in the invariant subspace that we are considering for the background geometry, as we have commented above. Moreover, it is remarkable that all the functions of $\bar{p}_j$ that appear in the background-dependent Hamiltonian factors of Eq. \eqref{eq: VB3} turn out to be non-negative. This fact allows us to address the only relevant ambiguity encountered to define those background factors, which arises when dealing with a product of noncommuting operators, a situation that can only occur when the operators belong to the same sector. To deal with this issue, we employ a symmetric algebraic ordering for products of (a power of) $\bar{\Omega}_j$ with a non-negative function of $\bar{p}_j$, representing it by (this power of) $\hat{\Omega}_j$ multiplied from the left and right by the square root of the operator corresponding to the considered function of $\bar{p}_j$. In this way, we attain an operator representation of the Hamiltonian constraint of the perturbed system in our LQC-Hybrid approach. The only task left is to find quantum solutions. To construct such solutions and derive with them master equations for the perturbations, one can follow a similar approach to that explained in Ref. \cite{LMM}. This will be the subject of future research. 

\section{Conclusions \label{sec: VII}}

We have developed a Hamiltonian formalism tailored to perturbed cosmological scenarios around a Kantowski-Sachs geometry, minimally coupled with a homogeneous scalar field. To achieve this aim, we first provided a brief outline of the Hamiltonian description of the background spacetime, addressing an important question, namely, how to perform a canonical transformation that adjusts the background variables to a new set, better suited for the subsequent quantization of the model. 

Having laid this groundwork, we then carried out a detailed examination of the possible perturbations of the background. In this process, a pivotal aspect is the expansion of the perturbations in spherical harmonics and Fourier modes, exploiting the spatial symmetries of the background. The consideration of a compact section ensures that zero modes can be isolated from such expansion, allowing us to separate them from the rest and treat them exactly, rather than as perturbations, and thus also avoid any possible infrared divergence in sums of contributions over all modes (the non-compact case with continuous Fourier modes should then be reachable in a suitable limit). For this expansion, we have employed real eigenfunctions of the Laplace-Beltrami operator associated with the spatial sections. In the decomposition we have introduced, the treatment of modes becomes much simpler, as it allows us to handle efficiently all the spatial dependence of the perturbations. The use of a real basis also facilitates the understanding of the physical nature of the modes, making their identification with phase space variables more straightforward. Leveraging the behavior of spherical harmonics under parity transformations, we have classified the perturbations into two classes, namely, axial and polar perturbations. Both of them have decoupled dynamical equations at leading perturbative order. This property allows us to analyze them separately. We have focused our attention primarily on the study of the axial perturbations, owing to the simplicity of their contribution compared to their polar counterpart. For these axial modes, the analysis has been limited to the leading perturbative order, corresponding in fact to a second-order perturbative truncation of the action of the system. This action, restricted to the axial sector, is formed by three types of terms. These are a presymplectic or Legendre term (i.e., the difference between the Lagrangian and the constraint terms), the perturbative constraints, and the perturbative contribution to the Hamiltonian constraint (which we also call the perturbative Hamiltonian). In addition, we divided the axial action into two different types of blocks: One for modes with $l\geq2$, and the other for modes with $l=1$, for which we have carried out a separate study adapted to their peculiarities. Let us also recall that, in the axial case, zero modes and the perturbation of the scalar field need not be considered. This is so because axial contributions vanish when $l=0$, and the scalar field admits only polar perturbations.

In the construction of the Hamiltonian formulation, a crucial step is the description of the perturbations by means of perturbative gauge invariants, perturbative constraints (which are also gauge invariants but must vanish) and momenta of these constraints (which are gauge variables). To achieve this description, we have had to introduce canonical transformations for the perturbations, which depend on both the background variables and the eigenvalues of the modes. Additionally, at a specific stage, we have found it necessary to redefine certain Lagrange multipliers of our theory, redefinitions that we have shown that are always valid at the order employed in our perturbative truncation. For the lapse function, we have chosen a densitization which is common in LQC and allows a direct relationship with previous results obtained for the background spacetime \cite{GBA}. 

For the modes with $l\geq2$, our approach leads to a diagonal perturbative Hamiltonian that depends quadratically on perturbative gauge invariants, but not in other perturbative variables. The background dependence of this Hamiltonian is captured in two mode-dependent factors. One of them represents a global factor, while the other accompanies the square of the configuration perturbative gauge invariants, and would play the role of a background-dependent frequency in the context of a harmonic oscillator. Moreover, these background factors are simple functions of the basic background variables employed in LQC. In the few cases in which their summands are not polynomials, they are rational functions that are positive for nonvanishing background densitized triads. Therefore, in an LQC-Hybrid approach, the operator representation of these factors with loops techniques does not pose any serious complication. We have also seen that the modes with $l=1$ are pure gauge, except for those with $n=0$. For these three modes (corresponding to the three possible values of $m$), an approach similar to that followed for the case $l\geq 2$ is applicable, resulting also in a diagonal perturbative Hamiltonian. 

The introduction of background-dependent canonical transformations for the perturbations might give the impression that we are disrupting the global canonical structure of the system. To dispel these doubts and show how the perturbed system remains canonical, we have added an appendix. At the perturbative order of our truncation, the main effect of the transformations on the background is the inclusion of modifications that are quadratic in the perturbations, in order to preserve the choice of a canonical set of variables. 
 
Starting with our Hamiltonian formulation, we have managed to discuss for the first time in the LQC literature the details of a hybrid quantization of the Kantowski-Sachs spacetime and its axial perturbations. Since the primary quantum geometry effects of the model are rooted in the background, we have embraced LQC to quantize the background variables, leveraging recent advancements in the loop quantization of Kantowski-Sachs \cite{GBA}. Simultaneously, we have adopted a Fock (or either a Schrödinger) representation for the quantization of the perturbative modes, acknowledging that perturbations can be treated using a more conventional representation, at least in certain regimes of direct interest. Nonetheless, let us comment that the hybrid quantization can also be adapted for combining other alternative quantum approaches, both for the background and the perturbations. With our hybrid strategy, we have been able to impose the perturbative constraints in a straightforward manner at the quantum level. They imply that physical states cannot depend on purely gauge modes of the perturbations. In parallel, we have carried out the quantization of the perturbative Hamiltonian in a Fock representation that is essentially unique, according to Ref. \cite{AT}, provided we require it to preserve the spatial background isometries and implement the dynamics of the perturbations as unitary transformations (when the background can be treated classically or effectively). This Hamiltonian is quadratic in the perturbative gauge invariant modes, with no mixing terms. It is important to highlight that, during the quantization process, we have not fixed the gauge freedom in any way.

The results of this work pave the way for interesting new projects. First, a natural continuation of this work is to complete the perturbative analysis by exploring polar perturbations. This continuation would mirror the procedure followed for axial perturbations, but with the added complexity that calculations would be more intricate, owing to the greater number of terms in the polar case compared to the axial contribution. The second line of research is analyzing in detail the hybrid quantization of the perturbations discussed in this paper. The goal of this study would be the inclusion of quantum corrections in the computation of measurable predictions for anisotropic cosmologies. This endeavor aligns with current efforts to investigate consequences of LQC for cosmological perturbations \cite{hyb-review,hyb1,AAN2,AAN3}. Finally, a more ambitious line of investigation involves contemplating the correspondence between Kantowski-Sachs and the interior of a Schwarzschild black hole, as we have commented before. The objective would be to extend the results about the quantum behavior of the interior and its perturbations to the exterior geometry of the black hole, obtaining in this way predictions about quantum effects, for instance, in greybody factors \cite{CS2} or other aspects of gravitational radiation. 

\acknowledgments

This work was supported by Project No. MICINN PID2020-118159GB-C41 from Spain.

\appendix

\section{CORRECTIONS OF THE BACKGROUND VARIABLES \label{sec: app}}  

In this appendix, we show how to recover a canonical structure for the system formed by the background and its perturbations when we pass to a perturbative gauge invariant description. As indicated in the main text, the transformations applied with this aim to the perturbations, obtained with a generating function that we generically call $\mathscr{A}$, spoil the global canonical structure when we consider the original background variables. Following an approach similar to that of Ref. \cite{LMM}, our goal is to redefine the background variables with perturbative corrections to restore the canonical structure. Let us first express the total action using an appropriate notation, motivated by Ref. \cite{LMM},
\begin{equation}
    S = \int_{\mathbb{R}}\text{d}t\left(\sum_a(w_q^a)_{,t}w_p^a + \frac{\epsilon^2}{\kappa}\sum_b\sum_{\mathfrak{n},\lambda}(X_{q_b}^{\mathfrak{n},\lambda})_{,t}X_{p_b}^{\mathfrak{n},\lambda} - \textbf{H}[w^a,X_b^{\mathfrak{n},\lambda};\Tilde{N},h_c^{\mathfrak{n},\lambda}]\right).
\end{equation}
In this equation, the old perturbative variables are called $\{X_b^{\mathfrak{n},\lambda}\} = \{X_{q_b}^{\mathfrak{n},\lambda}, X_{p_b}^{\mathfrak{n},\lambda}\}$, while the old background variables are $\{w^a\} = \{w^a_q, w^a_p\}$, where $a$ goes from $1$ to $2$, $b$ labels all the perturbative variables, and $q$ and $p$ subscripts stand for configuration and momenta variables. The total Hamiltonian $\textbf{H}$ is the sum of the background and the perturbative contributions. For the purposes of our discussion, let us express it as follows:
\begin{equation}
    \begin{aligned}
        \textbf{H}[w^a,X_b^{\mathfrak{n},\lambda};\Tilde{N},h_c^{\mathfrak{n},\lambda}] &= \Tilde{H}_{\text{KS}}[w^a;\Tilde{N}] + H_{\text{P}}[w^a,X_b^{\mathfrak{n},\lambda};\Tilde{N},h_c^{\mathfrak{n},\lambda}]\\
        &= \Tilde{H}_{\text{KS}}[w^a;\Tilde{N}] + \epsilon^2\sum_c\mathcal{C}^c_1[w^a,X_b^{\mathfrak{n},\lambda};h_c^{\mathfrak{n},\lambda}] + \epsilon^2\mathcal{H}_2[w^a,X_b^{\mathfrak{n},\lambda};\Tilde{N}].
    \end{aligned}
\end{equation}
Here, $\{\mathcal{C}^c_1\}$ represents the set of first-order perturbative constraints, $\{h_c^{\mathfrak{n},\lambda}\}$ are the corresponding perturbative Lagrange multipliers, where $c$ labels all perturbative constraints, and $\mathcal{H}_2$ is the second-order Hamiltonian for the perturbations. Redefining our background variables with second-order perturbative corrections, once the canonical transformations are completed, we want that terms in the action additional to the total Hamiltonian become
\begin{equation}
    \int_{\mathbb{R}}\text{d}t\left(\sum_a(\Bar{w}_q^a)_{,t}\Bar{w}_p^a + \frac{\epsilon^2}{\kappa}\sum_b\sum_{\mathfrak{n},\lambda}(V_{q_b}^{\mathfrak{n},\lambda})_{,t}V_{p_b}^{\mathfrak{n},\lambda}\right),
\end{equation}
where $\{\Bar{w}^a\} = \{\Bar{w}^a_q, \Bar{w}^a_p\}$ and $\{V_b^{\mathfrak{n},\lambda}\} = \{V_{q_b}^{\mathfrak{n},\lambda}, V_{p_b}^{\mathfrak{n},\lambda}\}$ denote the new canonical sets for the background and the perturbations, respectively. Actually, the linear relationship between the old and new perturbative variables allows us to write
\begin{equation}
    X_{b}^{\mathfrak{n},\lambda} = \sum_{b',\lambda'} \left[\frac{\partial X_{b}^{\mathfrak{n},\lambda}}{\partial V_{q_{b'}}^{\mathfrak{n},\lambda'}}V_{q_{b'}}^{\mathfrak{n},\lambda'} + \frac{\partial X_{b}^{\mathfrak{n},\lambda}}{\partial V_{p_{b'}}^{\mathfrak{n},\lambda'}}V_{p_{b'}}^{\mathfrak{n},\lambda'}\right].
\end{equation}
Consequently, at the perturbative order of our truncation in the action, the corrected background variables can be computed to be 
\begin{equation}
    w^a_{q} = \Bar{w}^a_{q} + \frac{\epsilon^2}{2\kappa}\sum_b\sum_{\mathfrak{n},\lambda} \left[X^{\mathfrak{n},\lambda}_{q_b}\frac{\partial X^{\mathfrak{n},\lambda}_{p_b}}{\partial \Bar{w}^a_{p}} - \frac{\partial X^{\mathfrak{n},\lambda}_{q_b}}{\partial \Bar{w}^a_{p}}X^{\mathfrak{n},\lambda}_{p_b}\right], \qquad
        w^a_{p} = \Bar{w}^a_{p} - \frac{\epsilon^2}{2\kappa}\sum_b\sum_{\mathfrak{n},\lambda} \left[X^{\mathfrak{n},\lambda}_{q_b}\frac{\partial X^{\mathfrak{n},\lambda}_{p_b}}{\partial \Bar{w}^a_{q}} - \frac{\partial X^{\mathfrak{n},\lambda}_{q_b}}{\partial \Bar{w}^a_{q}}X^{\mathfrak{n},\lambda}_{p_b}\right],
\end{equation}
where the original variables are understood as functions of the new ones when calculating the partial derivatives\footnote{Up to our perturbative order, the partial derivatives with respect to $\Bar{w}^a$ are the same as the partial derivatives with respect to $w^a$.}, which must be taken keeping the new perturbative variables fixed. Let us comment that these expressions differ from the ones given in Ref. \cite{LMM} by a sign in the corrections in order to align the result with our convention for the background Poisson bracket. Substituting this relation, we can verify that, at the perturbative order of our truncation,
\begin{equation}
    \int_{\mathbb{R}}\text{d}t\sum_a(w_q^a)_{,t}w_p^a  = \int_{\mathbb{R}}\text{d}t\sum_a(\Bar{w}_q^a)_{,t}\Bar{w}_p^a + \int_{\mathbb{R}}\text{d}t\sum_a\left[(w_p^a - \Bar{w}_p^a)(\Bar{w}_q^a)_{,t} - (w_q^a - \Bar{w}_q^a)(\Bar{w}_p^a)_{,t}\right].
\end{equation}
The last term of this formula can be expressed in a more familiar manner as
\begin{equation}
    \int_{\mathbb{R}}\text{d}t\sum_a\left[(w_p^a - \Bar{w}_p^a)(\Bar{w}_q^a)_{,t} - (w_q^a - \Bar{w}_q^a)(\Bar{w}_p^a)_{,t}\right] = \int_{\mathbb{R}}\text{d}t\frac{\epsilon^2}{\kappa}\sum_a\left[\frac{\partial \mathscr{A}}{\partial \Bar{w}_q^a}(\Bar{w}_q^a)_{,t} + \frac{\partial \mathscr{A}}{\partial \Bar{w}_p^a}(\Bar{w}_p^a)_{,t}\right].
\end{equation}
The outcome for the perturbative modes is already known, as we employed a similar expression in previous calculations,
\begin{equation}
    \int_{\mathbb{R}}\text{d}t\sum_b\sum_{\mathfrak{n},\lambda}(X_{q_b}^{\mathfrak{n},\lambda})_{,t}X_{p_b}^{\mathfrak{n},\lambda} = \int_{\mathbb{R}}\text{d}t\sum_b\sum_{\mathfrak{n},\lambda}(V_{q_b}^{\mathfrak{n},\lambda})_{,t}V_{p_b}^{\mathfrak{n},\lambda} - \int_{\mathbb{R}}\text{d}t\sum_a\left[\frac{\partial \mathscr{A}}{\partial \Bar{w}_q^a}(\Bar{w}_q^a)_{,t} + \frac{\partial \mathscr{A}}{\partial \Bar{w}_p^a}(\Bar{w}_p^a)_{,t}\right] + \int\text{d}\mathscr{A}.
\end{equation}
The last term is the integral of a total derivative, and can be ignored in the following discussion. On the other hand, at the considered perturbative order in the action, the total Hamiltonian can be rewritten using a Taylor expansion around the new background variables as
\begin{equation}
    \textbf{H}[w^a,X_b^{\mathfrak{n},\lambda};\Tilde{N},h_c^{\mathfrak{n},\lambda}] = \Tilde{H}_{\text{KS}}[\Bar{w}^a;\Tilde{N}] + \sum_{a'}(w^{a'} - \Bar{w}^{a'})\frac{\partial \Tilde{H}_{\text{KS}}}{\partial \Bar{w}^{a'}}[\Bar{w}^a;\Tilde{N}] + H_{\text{P}}\left[\Bar{w}^a,X_b^{\mathfrak{n},\lambda}[\Bar{w}^a,V_b^{\mathfrak{n},\lambda}];\Tilde{N},h_c^{\mathfrak{n},\lambda}\right],
\end{equation}
where the first and third terms correspond to the background Hamiltonian and the perturbative Hamiltonian, respectively, expressed in terms of the new sets of variables. As for the second term, it can be rewritten as
\begin{equation}
    \sum_{a}(w^a - \Bar{w}^a)\frac{\partial \Tilde{H}_{\text{KS}}}{\partial \Bar{w}^a} = \frac{\epsilon^2}{2\kappa}\left[\sum_{a}\left(\frac{\partial \mathscr{A}_{(3)}}{\partial \Bar{w}^a_q}\frac{\partial \Tilde{H}_{\text{KS}}}{\partial \Bar{w}^a_p} - \frac{\partial \Tilde{H}_{\text{KS}}}{\partial \Bar{w}^a_q}\frac{\partial \mathscr{A}_{(3)}}{\partial \Bar{w}^a_p}\right) + \sum_{a}\left(\frac{\partial \mathscr{A}_{(1)}}{\partial \Bar{w}^a_q}\frac{\partial \Tilde{H}_{\text{KS}}}{\partial \Bar{w}^a_p} - \frac{\partial \Tilde{H}_{\text{KS}}}{\partial \Bar{w}^a_q}\frac{\partial \mathscr{A}_{(1)}}{\partial \Bar{w}^a_p}\right)\right],
\end{equation}
where $\mathscr{A}_{(1)}$ and $\mathscr{A}_{(3)}$ are, respectively, the generating functions of type $1$ and $3$ associated with the canonical transformation for the perturbations. Thus, altogether, the final form of the total Hamiltonian of the perturbed system is given, at the considered perturbative order (and ignoring for simplicity redefinitions of Lagrange multipliers), by
\begin{equation}
    \textbf{H}[w^a,X_b^{\mathfrak{n},\lambda};\Tilde{N},h_c^{\mathfrak{n},\lambda}] = \Tilde{H}_{\text{KS}}[\Bar{w}^a;\Tilde{N}] + \epsilon^2\sum_c\mathcal{C}^c_1[\Bar{w}^a,V_b^{\mathfrak{n},\lambda};h_c^{\mathfrak{n},\lambda}] + \epsilon^2\left.\left[\mathcal{H}_2 + \frac{1}{\kappa}\{\mathscr{A},\Tilde{H}_{\text{KS}}\}_{\text{B}}\right]\right|_{[\Bar{w}^a,V_b^{\mathfrak{n},\lambda};\Tilde{N}]}.
\end{equation}
The interpretation of this result is straightforward. The only modification required in our expressions is the substitution of $w^a$ for $\Bar{w}^a$, which can be accomplished without complication, thereby preserving the validity of the entire discussion concerning the axial perturbations. It is worth noticing that no equations of motion were necessary during our derivation, since Poisson brackets are defined as a bilinear map to functions on phase space.


\begin{thebibliography}{299}

\bibitem{ringdown} K.D. Kokkotas and B.G. Schmidt, Quasi-normal modes of stars and black holes, Living Rev. Relativity {\bf 2}, 2 (1999). 
\bibitem{Bardeen} J.M. Bardeen, Gauge-invariant cosmological perturbations, Phys. Rev. D {\bf 22}, 1882 (1983).
\bibitem{Mukhanov} V. Mukhanov, Quantum theory of gauge invariant cosmological perturbations, Zh. Eksp. Teor. Fiz. {\bf 94}, 1 (1988) [Sov. Phys. JETP {\bf 67}, 1297 (1988)].
\bibitem{Sasaki} M. Sasaki, Gauge invariant scalar perturbations in the new inflationary universe, Prog. Theor. Phys. {\bf 70}, 394 (1983).
\bibitem{RW} T. Regge and J.A. Wheeler, Stability of a Schwarzschild singularity, Phys. Rev. {\bf 108}, 1063 (1957).
\bibitem{M} V. Moncrief, Gravitational perturbations of spherically symmetric systems. I. The exterior problem, Ann. Phys. (N.Y.) {\bf 88}, 323 (1974).
\bibitem{A&L} A. Ashtekar and J. Lewandowski, Background independent quantum gravity: A status report, Classical Quantum Gravity {\bf 21}, R53 (2004).
\bibitem{Thiemann} T. Thiemann, \textit{Modern Canonical Quantum General Relativity} (Cambridge University Press, Cambridge, UK, 2007).
\bibitem{A&S} A. Ashtekar and P. Singh, Loop quantum cosmology: A status report, Classical Quantum Gravity {\bf 28}, 213001 (2011).
\bibitem{GMM} G.A. Mena Marug\'an, A brief introduction to loop quantum cosmology, AIP Conf. Proc. {\bf 1130}, 89 (2009).
\bibitem{APS1} A. Ashtekar, T. Paw\l{}owski, and P. Singh, Quantum nature of the big bang, Phys. Rev. Lett. {\bf 96}, 141301 (2006).
\bibitem{APS2} A. Ashtekar, T. Paw\l{}owski, and P. Singh, Quantum nature of the big bang: Improved dynamics, Phys. Rev. D. {\bf 74}, 084003 (2006).
\bibitem{K&S} R. Kantowski and R.K. Sachs, Some spatially inhomogeneous dust models, J. Math. Phys. (N.Y.) {\bf 7}, 443
(1966).
\bibitem{E.Weber} E. Weber, Kantowski-Sachs cosmological models as big-bang models, J. Math. Phys. (N.Y.) {\bf 26}, 1308 (1984).
\bibitem{A&B} A. Ashtekar and M. Bojowald, Quantum geometry and the Schwarzschild singularity, Classical Quantum Gravity {\bf 23}, 391 (2005).
\bibitem{LBH1} L. Modesto, Loop quantum black hole, Classical Quantum Gravity {\bf 23}, 5587 (2006).
\bibitem{LBH2} D. Cartin and G. Khanna, Wave functions for the Schwarzschild black hole interior, Phys. Rev. D {\bf 73}, 104009 (2006).
\bibitem{B&V} C.G. B\"{o}hmer and K. Vandersloot, Loop quantum dynamics of Schwarzschild interior, Phys. Rev. D {\bf 76}, 104030 (2007).
\bibitem{Chiou} D.W. Chiou, Phenomenological loop quantum geometry of the Schwarzschild black hole, Phys. Rev. D {\bf 78}, 064040 (2008).
\bibitem{LBH3} M. Campiglia, R. Gambini, and J. Pullin, Loop quantization of spherically symmetric midisuperspaces: The interior problem, AIP Conf. Proc. {\bf 977}, 52 (2008).
\bibitem{LBH4} H.M. Haggard and C. Rovelli, Quantum-gravity effects outside the horizon spark black to white hole tunnelling, Phys. Rev. D {\bf 92}, 104020 (2015).
\bibitem{LBH5} A. Corichi and P. Singh, Loop quantization of the Schwarzschild interior revisited, Classical Quantum Gravity {\bf 33}, 055006 (2016).
\bibitem{JPS} J. Olmedo, S. Saini, and P. Singh, From black hole to white holes: A quantum gravitational symmetric bounce, Classical Quantum Gravity {\bf 34}, 225011 (2017).
\bibitem{LBH6} J. Cortez, W. Cuervo, H.A. Morales-Técotl, and J.C. Ruelas, On effective loop quantum geometry of Schwarzschild interior, Phys. Rev. D {\bf 95}, 064041 (2017).
\bibitem{LBH7} E. Bianchi, M. Christodoulou, F. D’Ambrosio, H.M. Haggard, and C. Rovelli, White holes as remnants: A surprising scenario for the end of a black hole, Classical Quantum Gravity {\bf 35}, 225003 (2018).
\bibitem{LBH8} E. Alesci, S. Bahrami, and D. Pranzetti, Quantum gravity predictions for black hole interior geometry, Phys. Lett. B {\bf 797}, 134908 (2019).
\bibitem{LBH9} M. Bojowald, Black-hole models in loop quantum gravity, Universe {\bf 6}, 125 (2020).
\bibitem{LBH10} R. Gambini, J. Olmedo, and J. Pullin, Spherically symmetric loop quantum gravity: Analysis of improved dynamics, Classical Quantum Gravity {\bf 37}, 205012 (2020).
\bibitem{LBH11} J.G. Kelly, R. Santacruz, and E. Wilson-Ewing, Effective loop quantum gravity framework for vacuum spherically symmetric spacetimes, Phys. Rev. D {\bf 102}, 106024 (2020).
\bibitem{LBH12} J.G. Kelly, R. Santacruz, and E. Wilson-Ewing, Black hole collapse and bounce in effective loop quantum gravity, Classical Quantum Gravity {\bf 38}, 04LT01 (2021).
\bibitem{AOS} A. Ashtekar, J. Olmedo, and P. Singh, Quantum transfiguration of Kruskal black holes, Phys. Rev. Lett. {\bf 121}, 241301 (2018).
\bibitem{AOS2} A. Ashtekar, J. Olmedo, and P. Singh, Quantum extension of the Kruskal spacetime, Phys. Rev. D {\bf 98}, 126003 (2018).
\bibitem{AO} A. Ashtekar and J. Olmedo, Properties of a recent quantum extension of the Kruskal geometry, Int. J. Mod. Phys. D {\bf 29}, 2050076 (2020).
\bibitem{BH_Con} C. Zhang, Y. Ma, S. Song, and X. Zhang, Loop quantum Schwarzschild interior and black hole remnant, Phys. Rev. D {\bf 102}, 041502(R) (2020).
\bibitem{GQM2} A. García-Quismondo and G.A. Mena Marugán, Two-time alternative to the Ashtekar-Olmedo-Singh black hole interior, Phys. Rev. D {\bf 106}, 023532 (2022).
\bibitem{BH_GAB} B. Elizaga Navascu\'es, A. Garc\'{\i}a-Quismondo, and G.A. Mena Marug\'an, Hamiltonian formulation and loop quantization of a recent extension of the Kruskal spacetime, Phys. Rev. D {\bf 106}, 043531 (2022).
\bibitem{BH_Cong} C. Zhang, Y. Ma, S. Song, and X. Zhang, Loop quantum deparametrized Schwarzschild interior and discrete black hole mass, Phys. Rev. D {\bf 105}, 024069 (2022).
\bibitem{GBA} B. Elizaga Navacues, G.A. Mena Marugán, and A. Mínguez-Sánchez, Extended phase space quantization of a black hole interior model in loop quantum cosmology, Phys. Rev. D {\bf 108}, 106001 (2023).
\bibitem{Lan} D. Langlois, Hamiltonian formalism and gauge invariance for linear perturbations in inflation, Classical Quantum Gravity {\bf 11}, 389 (1994).
\bibitem{DC} D. Brizuela, High-order perturbation theory of spherical spacetimes with application to vacuum and perfect fluid matter, http://hdl.handle.net/10486/121, Ph.D thesis supervised by G.A. Mena Marugán and J.M. Mart\'{\i}n-Garc\'{\i}a, Univ. Aut\'onoma de Madrid, Madrid, 2011.
\bibitem{LMM} L. Castell\'o Gomar, M. Martín-Benito, and G.A. Mena Marugán, Gauge-invariant perturbations in hybrid quantum cosmology, J. Cosmol. Astropart. Phys. {\bf 06} (2015) 045.
\bibitem{AT} J. Cortez, B. Elizaga Navacu\'es, G.A. Mena Marugán, A. Torres-Caballeros, and J.M. Velhinho, Fock quantization of a Klein-Gordon field in the interior geometry of a nonrotating black hole, Mathematics {\bf 11}, 3922 (2023).
\bibitem{hyb-review} B. Elizaga Navascu\'es and G.A. Mena Marug\'an,  Hybrid loop quantum cosmology: An overview, Front. Astron. Space Sci. {\bf 8}, 624824 (2021).
\bibitem{FMO} M. Fern\'andez-M\'endez, G.A. Mena Marug\'an, and J. Olmedo, Hybrid quantization of an inflationary universe, Phys. Rev. D {\bf 86}, 024003 (2012).
\bibitem{hyb1} L. Castell\'o Gomar, G.A. Mena Marug\'an, D. Mart\'{\i}n de Blas, and J. Olmedo, Hybrid loop quantum cosmology and predictions for the Cosmic Microwave Background, Phys. Rev. D {\bf 96}, 103528 (2017).
\bibitem{hyb2} B. Elizaga Navascu\'es, D. Mart\'{\i}n de Blas, and G.A. Mena Marug\'an, The vacuum state of primordial fluctuations in hybrid loop quantum cosmology, Universe {\bf 4}, 98 (2018).
\bibitem{AAN} I. Agullo, A. Ashtekar, and W. Nelson, A quantum gravity extension of the inflationary scenario, Phys. Rev. Lett. {\bf 109}, 251301 (2012).
\bibitem{AAN2} I. Agullo, A. Ashtekar, and W. Nelson, Extension of the quantum theory of cosmological perturbations to the Planck era, Phys. Rev. D {\bf 87}, 043507 (2013).
\bibitem{AAN3} I. Agullo, A. Ashtekar, and W. Nelson, The pre-inflationary dynamics of loop quantum cosmology: Confronting quantum gravity with observations, Classical Quantum Gravity {\bf 30}, 085014 (2013).
\bibitem{DB2} D. Brizuela, J.M. Martín-García, and G.A. Mena Marugán, High-order gauge-invariant perturbations of a spherical spacetime, Phys. Rev. D {\bf 76}, 024004 (2007).
\bibitem{DB3} D. Brizuela, and J.M. Martín-García, Hamiltonian theory for the axial perturbations of a dynamical spherical background, Classical Quantum Gravity {\bf 26}, 015003 (2008).
\bibitem{DB1} D. Brizuela, J.M. Martín-García, and G.A. Mena Marugán, Second and higher-order perturbations of a spherical spacetime, Phys. Rev. D {\bf 74}, 044039 (2006).
\bibitem{CS1} C.F. Sopuerta and M. Lenzi, Master functions and equations for perturbations of vacuum spherically symmetric spacetimes, Phys. Rev. D {\bf 104}, 084053 (2021).  
\bibitem{MMO} M. Mart\'{\i}n-Benito, G.A. Mena Marug\'an, and J. Olmedo, Further improvements in the understanding of isotropic loop quantum cosmology, Phys. Rev. D {\bf 80}, 104015 (2009).
\bibitem{Prescrip_G} G.A. Mena Marug\'an, J. Olmedo, and T. Paw\l{}owski, Prescriptions in loop quantum cosmology: A comparative analysis, Phys. Rev. D {\bf 84}, 064012 (2011).
\bibitem{CS2} C.F. Sopuerta and M. Lenzi, Black hole greybody factors from Korteweg\textendash{}de Vries integrals: Theory, Phys. Rev. D {\bf 107}, 044010 (2023).

\end{thebibliography}
\end{document}